\documentclass[traditabstract]{aa}
\usepackage{psfrag,graphicx}
\usepackage{color}
\usepackage{txfonts}
\usepackage{natbib}
\usepackage{breqn}
\usepackage[T1]{fontenc}
\usepackage{ae,aecompl}

\title{ Magnetic fields in primordial accretion disks}
\titlerunning{Magnetic fields during the formation of Pop. III stars}

\author{M.~A.~Latif \inst{1,2}
\and
D.~R.~G.~Schleicher \inst{3}
}
\institute{Sorbonne Universités, UPMC Univ Paris 06, UMR 7095, Institut d'Astrophysique de Paris, F-75014, Paris, France
\and
CNRS, UMR 7095, Institut d'Astrophysique de Paris, F-75014, Paris, France
\and
Departamento de Astronomía, Facultad Ciencias Físicas y Matemáticas, Universidad de Concepción,  Av. Esteban Iturra s/n Barrio Universitario, Casilla 160-C, Chile
}

\authorrunning{Latif \& Schleicher}

\begin{document}

\bibliographystyle{aa}


\begin{abstract}
{
Magnetic fields are considered as a vital ingredient of contemporary star formation, and may have been important during the formation of the first stars in the presence of an efficient amplification mechanism. Initial seed fields are provided via plasma fluctuations, and are subsequently amplified by the small-scale dynamo, leading to a strong tangled magnetic field. Here we explore how the magnetic field provided by the small-scale dynamo is further amplified via the $\alpha-\Omega$ dynamo in a protostellar disk and assess its implications. For this purpose, we consider two characteristic cases, a typical Pop.~III star with $10$~M$_\odot$ and an accretion rate of $10^{-3}$~M$_\odot$~yr$^{-1}$, and a supermassive star with $10^5$~M$_\odot$ and an accretion rate of $10^{-1}$~M$_\odot$~yr$^{-1}$. For the $10$~M$_\odot$ Pop.~III star, we find that coherent magnetic fields can be produced on scales of at least $100$~AU, which are sufficient to drive a jet with a luminosity of $100$~L$_\odot$ and a mass outflow rate of $10^{-3.7}$~M$_\odot$~yr$^{-1}$. For the supermassive star, the dynamical timescales in its environment are even shorter, implying smaller orbital timescales and an efficient magnetization out to at least $1000$~AU. The jet luminosity corresponds to $\sim10^{6.0}$~L$_\odot$, and a mass outflow rate of  $10^{-2.1}$~M$_\odot$~yr$^{-1}$. We expect that the feedback from the supermassive star can have a relevant impact on its host galaxy.}
\end{abstract}

\maketitle
\section{Introduction}
Magnetic fields play an important role in contemporary star formation. They  strongly contribute to the transfer of angular momentum \citep{Hawley91, Fromang04, Shu07, Hennebelle08}, launch collimated jets and outflows \citep{Shu95, Banerjee06, Fendt09, Wang10},  stabilize the accretion disk and help in suppressing fragmentation \citep{Hennebelle08b, Banerjee09, Price08}. 

In studies of primordial star formation, it was originally less clear whether the role of magnetic fields has  been significant, as the initial magnetic field strength was very uncertain within the framework of $\Lambda$CDM \citep[see e.g.][]{Grasso01,Kulsrud2008,Ryu2012, Widrow12}. The first cosmological simulations had shown that primordial stars form in dark matter halos with $10^5-10^6$~M$_\odot$ at $z\sim20$ \citep{Abel2002,Bromm2002} or even earlier \citep{Reed2005,Gao2010}. Subsequently, significant efforts were dedicated to a more realistic modeling and to investigate the longer-term evolution of these systems.  \citet{Yoshida08} performed the first simulations revealing the formation of a self-gravitating disk from cosmological initial conditions, while \citet{Clark11} and \citet{Greif12} explored the fragmentation of the resulting disks. These predominantly Lagrangian-type simulations have been complemented with grid-based simulations using the adaptive mesh refinement (AMR) technique, confirming the formation of disks and the potential fragmentation \citep{LatifPopIII13, BovinoHD2014}. \citet{Inayoshi14},\citet{Tanaka14} and \citet{Latif2014Disk,Latif2015c} have employed analytical models to further investigate the long term evolution of clumps forming as a consequence of disk fragmentation. These studies suggest that the clumps move inward due to the short migration time scale and merge with the central object.

As we discuss below, magnetic fields may  have played a relevant role during primordial star formation due to the efficient amplification of magnetic fields. Observationally, there is no evidence that magnetic fields were present during Pop.~III star formation, but there is a number of independent observations suggesting the existence of magnetic fields at rather early stages in the Universe. The far-infrared - radio correlation, a correlation between star formation activity and synchrotron emission due to magnetic fields \citep{Yun01}, has been demonstrated to hold at least until $z\sim2$ via Herschel results \citep{Ivison10, Magnelli15}, implying strong magnetic fields at a time when the Universe was about $3$~billion years old. Similarly, investigations exploring the correlation between quasar rotation measures and the number of normal galaxies along the line of sight have indicated that Milky-Way type magnetic fields are present in galaxies at  $z\sim2$ \citep{Kronberg08, Bernet08}. In quasars, strong rotation measures have been observed even at $z=5$, showing that strong magnetic fields can build up at early stages in the central regions of active galactic nuclei \citep{2012arXiv1209.1438H}. Even in the intergalactic medium, blazar observations  point at a minimum magnetic field strength of   $ \gtrsim 10^{-18}$~G \citep{Dermer2011,Tavecchio11, Dolag11,ARLEN2014}.

While the strength of a primordial field is quite uncertain, astrophysical mechanisms suggest an initial field strength of $\rm \sim10^{-20}$~G.  The Biermann battery operates on kpc scales under favourable conditions \citep{Biermann1950}, while both the Weibel instability \citep{Weibel1959}  and the generation of spontaneously emitted magnetic fields occurs on much smaller scales, comparable to a few Debye lengths. Recent calculations by \citet{Schlickeiser12} suggest the production of $\sim10^{-10}$~G magnetic fields in protogalaxies due to plasma fluctuations \citep{Yoon14}. So, the latter two mechanisms can produce a magnetic field on small scales.  The small-scale dynamo \citep{Kazantsev68} can enhance the magnetic field strength via turbulence \citep{Brandenburg05, Schoberb} and lead to the rapid production of a strong tangled magnetic field. After saturation on small-scales, the dynamo will enter the non-linear regime, and larger scales will be magnetised on their respective eddy-turnover timescales \citep{Scheko02,Schleicher13a}. The conditions for the dynamo to operate require turbulence and a critical magnetic Reynolds number ${\rm Rm_{crit} = v \ell}/ \eta$, where v and $\ell$ are the characteristic turbulent velocity and length scales and $\eta$ the turbulent diffusivity.  The minimum magnetic Reynolds number for the dynamo  to operate is about $\rm Rm_{crit} \sim $ 100, for details see \cite{Brandenburg05} and \cite{Schoberb}. Both the Ohmic and ambipolar resistivity during primordial star formation are found to be sufficiently low to stop efficient magnetic field amplification \citep{Schobera}. The amplification time corresponds to the timescale of turbulent fluctuations, and is therefore much shorter than the evolutionary timescale of the system.  

Cosmological simulations show that  turbulence is  driven by gravitational instabilites during the gravitational collapse in primordial halos forming at z=15-30 \citep{Turk2012,Latif2013a,Latif2013d}, as also expected from analytical considerations \citep{Elmegreen10, Klessen10}. Continuous infall and accretion onto the primordial disk further helps to drive turbulence even in the absence of other mechanisms like the magneto-rotational instability \citep{Hawley91, Balbus1998}.  The other requirement for  the small scale dynamo to operate is the coupling between the gas and the magnetic field. Magnetic dissipiative processes such as the Ohmic dissipation and the ambiploar diffusion are  less effective in primordial gas due to the higher temperatures and a larger ionization degree. \citet{Maki04,Maki07} and \citet{Susa15} have investigated the dissipation of  magnetic fields in primordial gas clouds and found that the resistivity is a few orders of magnitude lower than in present-day star forming clouds and the coupling of magnetic fields  with  the gas is very efficient. The latter ensures an important pre-condition for the operation of the dynamo. We note that similar results for the ionization degree have been obtained in an independent study by \citet{Glover09}.

Analytical studies suggest that the small-scale dynamo operates soon after virialization of the halo and leads to a strong magnetic field in approximate equipartition with turbulent energy \citep{Souza10, Schleicher2010Ma, Schobera}. The latter is rather independent of the initial seed field, as the initial amplification occurs on particularly small scales with rapid amplification. While it is difficult to resolve the dynamo process in numerical simulations during gravitational collapse, requiring a typical resolution of at least $64$ cells per Jeans length, the latter has been achieved in a number of studies during the past years \citep{Sur2010, Federrath11a, Peters12, Sur12, Turk2012, Latif13}. \cite{Turk2012} and \cite{Latif13} performed cosmological magneto-hydrodynamical simulations and confirmed  that the small-scale dynamo is operational during the formation of protogalaxies, and the operation of the small-scale dynamo has been confirmed even for turbulence driven by supernova explosions \citep{Balsara2004, Balsara05}.  Furthermore, substantial progress has  been made in the theoretical understanding of the dynamo, including the regime at high Mach numbers, different types of turbulence and a large range of magnetic Prandtl numbers \citep{Federrath11b, Schoberb, Bovino13, Schleicher13, Federrath14}.

While the fields produced by the small scale dynamo are strongly tangled, observations show coherent fields even on  galactic scales, even though perturbations may be significant \citep[e.g.][]{Beck96}. A large-scale field can be generated via the $\alpha$-$\Omega$ dynamo  in differentially rotating disks \citep{Steeback1966,Vainshtein1971,Parker1971, Brandenburg05}. Simulations of isolated disk galaxies have shown that the $\alpha$-$\Omega$ dynamo efficiently amplifies magnetic fields on their orbital timescale and  the magnetic field strength reaches equipartition value \citep{Gressel2008, WangAbel2009,Pakmor2013}. They even become dynamically important, launch outflows and constrain the star formation efficiency. Various studies \citep{Ruzmaikin1988,Schmitt1990,Kulsrud1999,Moss2012,Moss2013} suggest that the large scale dynamo is essential to generate ordered magnetic fields observed in spiral galaxies. We therefore emphasize that such dynamo processes have not only been confirmed in idealized scenarios, but in simulations of real astrophysical systems.

The turbulent amplification of magnetic fields in primordial disks has been studied by  \citet{Pudritz89} and \citet{Tan2004}. They  suggested the formation of coherent fields in high-redshift protogalactic disks via the $\alpha$-$\Omega$ dynamo, while \citet{Silk2006} explored the implications of the magneto-rotational instability as a potential magnetization mechanism. The formation of jets has been explored in simulations by \citet{Machida2006, Machida08} and \citet{Machida2013} starting from an initially coherent field. We  note that these studies typically explored the early phases of protostar formation and could not follow the evolution at later times, where the impact of magnetic fields is potentially more significant. Performing cosmological simulations which self-consistently include all physical processes still appears to exceed the current computational capabilities, in particular for studies aiming to investigate the long-term evolution. It is therefore necessary to employ analytical models  incorporating the most relevant processes which have been tested with individual simulations, such as the small-scale dynamo and the $\alpha$-$\Omega$ dynamo in the context of present work.

In this paper, we explore the formation of a large-scale magnetic field in the protostellar disk  during the formation of a Pop III star as well as a supermassive star and study its implications. For this purpose, we combine the $\alpha$-$\Omega$ dynamo  model of magnetic field amplification  with  recently developed disk models for the environment of Pop.~III stars and supermassive stars by \citet{Latif2015c,Latif2014Disk}. Assuming that the initial magnetization is provided by the small-scale dynamo, we calculate the required number of e-folding times to reach saturation with the $\alpha$-$\Omega$ dynamo. We further employ the resulting model of the magnetic field structure to estimate the luminosity and mass outflow rate of a jet, and discuss its implications for early structure formation.

This article is  structured as follows: In section \ref{diskstruct}, we present the general disk model and lay out the adopted framework to model magnetic field amplification and the potential of the field to drive a jet. In section \ref{results}, we apply the disk model both to a $10$~M$_\odot$ Pop. III star as well as a supermassive star, explore the resulting strength of the magnetic field and the potential power of the jet. We discuss our results and conclusions in section~\ref{discussion}.

\section{ Disk structure and Magnetic fields}\label{diskstruct}

\subsection{Generic disk model}

To compute the properties of a self-gravitating accretion disk formed in a primordial halo at $z \sim 20$, we employ the analytical framework  described by \citet{Latif2014Disk}. The  formation of such a disk is a consequence of angular momentum conservation in the aftermath of halo collapse which occurs in about a free-fall time \citep{Clark11,Greif12,LatifPopIII13,LatifVolonteri2015}. We presume that the disk is in an approximate steady-state  and is marginally stable due to self-regulation \citep[e.g.][]{Lodato2007}. The disk stability is described via the Toomre  criterion  \citep{Toomre1964}, given as
\begin{equation}
 Q = \frac{c_{s} \Omega}{\pi G \Sigma} .
\label{eq1}
\end{equation}
Here $\Omega$ and $c_s$ are the orbital frequency and the sound speed of the disk, $G$ is the gravitational constant and $\Sigma$ is the surface density of the disk. Under stationary conditions, the latter follows from the accretion rate $\dot{M}_{\rm tot}$ and the disk viscosity $\nu$ via
\begin{equation}
 \Sigma = \frac{\dot{M}_{tot}}{3 \pi \nu} .
\label{eq2}
\end{equation}
The disk scale-height can be estimated as $\rm H = \frac{c_s}{\Omega}$ assuming hydrostatic equilibrium in the vertical direction . This assumption is accurate for $Q\geq1$, while modifications are necessary once self-gravity becomes important in the vertical direction \citep{Lodato2007}.  

Assuming thermal equilibrium, the temperature of the disk is determined from the balance of heating and cooling. The viscous heating rate for a self-gravitating disk can be estimated by solving Eq.~(50) given in \cite{Lodato2007}. The heating rate per surface area is $\ Q_+ = \nu \Sigma (R \Omega')^2$. In case of Keplarian rotation, we have $\Omega_K = \sqrt{\frac{GM_{*}}{R^3}}$.  We  assume here that the disk is not fully Keplerian, as it should be partly supported by turbulent and thermal pressure. Therefore, we consider an efficiency parameter $\epsilon_K=\Omega/\Omega_K\sim0.5$, where $\Omega$ is the angular velocity of the disk. We here assume that the turbulent velocity is 0.5 times  the sound speed, which implies that the turbulent support is a fixed fraction of the thermal support. As the velocity profile approximately follows a Keplerian relation (even though with a slightly different slope and a potentially lower normalization), we expect an approximate scaling of $\epsilon_K$ as $R^{-1/2}$. As we show below, the main contribution to jet formation results from the inner parts of the disk, and neglecting this dependence is thus not too relevant.

The viscous heating rate for such a disk is given as
\begin{equation}
Q_+ = \frac{9}{4} \nu \Sigma \Omega^2 .
\label{eq3}
\end{equation}
 The cooling rate per surface area of the disk can be computed as follows: 
\begin{equation}
 Q_- = 2 H \Lambda_{\rm H/H_{2}} .
\label{eq4}
\end{equation}
Depending on the regime under consideration, $\Lambda_{\rm H/H_{2}}$ denotes the atomic or molecular cooling rate  in units of $\rm erg/cm^3/s$. For a disk around a typical Pop.~III star, the cooling on larger scales is driven by molecular hydrogen line cooling, while it is dominated by collisionally-induced emission (CIE) of molecular hydrogen in the interior. The detailed expressions for the cooling rates, including approximations employed for the optical depth, are given by \citet{Latif2014Disk}. We also note that the model was further extended by \citet{Latif2015c} for the disks of supermassive stars exposed to larger accretion rates, as the disks on large radii cool via molecular hydrogen line cooling, while the interior is dominated by atomic hydrogen line cooling as a result of the viscous heating.

The disk viscosity is parametrized via the $\alpha$-formalism of \cite{Shakura73} as
\begin{equation}
 \nu = \alpha_{vis} c_s H ,  
\label{eq5}
\end{equation} 
where $\alpha_{vis}$ is the viscous parameter. For Q=1, the latter can be determined using the above equations as:
\begin{equation}
\alpha_{vis} = \frac{\dot{M}_{tot} G}{ 3 c_s^3} .
\label{eq6}
\end{equation}
In the same way, the surface density of the disk follows as
\begin{equation}
 \Sigma = \frac{c_s \epsilon_K}{\pi G}  \sqrt{\frac{GM_{*}}{R^3}} ,
\label{eq7}
\end{equation} 
and the number density inside the disk is approximated as
\begin{equation}
n = \frac{\Sigma \epsilon_K}{2\pi \mu  m_p c_s}  \sqrt{\frac{GM_{*}}{R^3}} ,
\label{eq8}
\end{equation}
with $\mu$ the mean molecular weight. We adopt $\mu\sim2$ for the molecular gas and $\mu\sim1.2$ in the atomic cooling regime.

\subsection{From small-scale to large-scale magnetic fields}\label{magnetic}
In the following, we consider the evolution of magnetic fields for the disk model given above. For this purpose, we employ the dynamo model of \citet{Arshakian2009} for disk galaxies, where the initial magnetization is provided via the small-scale dynamo, and the $\alpha-\Omega$ dynamo subsequently produces an ordered magnetic field. The main parameters for the dynamo are the angular velocity $\Omega$, the disk height $H$, the disk radius $R$, and the mass density $\rho$, which are determined by the model above. The large-scale magnetic field component is defined via a volume integration over a characteristic scale $\lambda_{av} \sim 2R$ based on the Reynolds averaging procedure \citep{KrauseBook1980}. The small scale component then describes the deviation from the Reynolds average. For the largest eddies, we expect a characteristic scale comparable to the disk height $H$. For primordial disks with $H/R \sim 0.1-0.5$, the latter implies a characteristic ratio of $\lambda_{av} / H \sim 4-20$. For very thick disks, the assumption of scale separation may then only hold at an approximate level. However, in the presence of central massive objects, we expect that a relevant flattening will occur due to the gravity of the central source.

We assume here that well before the formation of the disk, a seed magnetic field has been generated for instance by the Biermann battery \citep{Biermann1950} or as a result of plasma fluctuations \citep{Schlickeiser12}, which was subsequently amplified via the small-scale dynamo \citep{Kazantsev68}. The latter starts in the kinematic phase, where amplification occurs on the viscous length scale $l_\nu$, with a characteristic timescale corresponding to the eddy-turnover time on that scale. Once the magnetic field has saturated on the viscous scale $l_\nu$, amplification continues on larger scales, leading to a linear growth in this regime. On the driving scale of turbulence, the magnetic field is then saturated after a few eddy-turnover times \citep{Scheko02, Beresnyak12, Schleicher13a}.  The equipartition field strength $B_*$ is defined as
\begin{equation}
 B_{*}^2/(8 \pi) = 0.5 \rho v_{t}^2 .
\label{eq16}
\end{equation}
Due to the short timescales involved, we expect this process to be completed when the disk has formed. We then estimate the magnetic field strength $B_{ss}$ produced by the small-scaled dynamo as
\begin{equation}
 B_{ss}^{2}/(8 \pi) = \epsilon B_{*} ,
\label{eq14}
\end{equation}
where $\epsilon$ is the saturation fraction and $v_t$ the turbulent velocity. We adopt a characteristic value of $\epsilon\sim0.3$, which is consistent with turbulence in the subsonic to transsonic regime \citep{Federrath11b}. Indeed, the turbulent velocity in the disk is typically of the order of the sound speed, so we adopt $v_t \sim 0.5 c_s$ \citep{Clark11, Greif12, Latif2013c, Latif2014Magnetic}.  We assume here that the turbulence is predominantly driven by infall and accretion onto the disk accompanied by gravitational instabilities \citep[c.f.][]{Elmegreen10, Klessen10}, but additional contributions may come from the magneto-rotational instability. The largest eddies within the disk have a maximum size $l_t$ given by the disk height $H$, which therefore defines the characteristic length scale of the magnetic field. Further, a small scale random field on the scales of the disk height is expected to induce a weak large-scale component. The strength of this component scales as $N^{-1/2}$, where $N = \frac{2 \pi R^2 H}{H^3}$ is the number of turbulent cells in the disk \citep{Hogan1983,Ruzmaikin1988,Arshakian2009}. Therefore, we expect an initial field strength of 
\begin{equation}
B_{init} = B_{ss}\left(\frac{H}{\sqrt{2 \pi} R} \right) 
\label{eq15}
\end{equation} 
on large scales. The latter provides the initial magnetic field strength that is subsequently amplified via the $\alpha-\Omega$ dynamo. To model the $\alpha$ effect, we adopt the parametrization given in Eq.  9.60 of \cite{KrauseBook1980}. We note that the helicity defining $\alpha$ is considered as an average over the disk height $H$, corresponding to the maximum scale of turbulent fluctuations. Thus, 
\begin{equation}
 \alpha =  \frac{ v_t^2 t_{cor}^2 \Omega}{H} ,
\label{eq01}
\end{equation}
where $t_{cor}$ is the correlation time scale defined as
\begin{equation}
t_{cor} =  \frac{H}{v_t} .
\label{eq02}
\end{equation}
 Combining the above equations, $\alpha$ can be computed as
 \begin{equation}
 \alpha =   \Omega H .
\label{eq9}
\end{equation}
The turbulent diffusivity  is calculated based on mixing length theory as
\begin{equation}
 \beta =  \frac{l_t v_t}{ 3} \sim \frac{Hc_s}{3} .
\label{eq10}
\end{equation}
The joint action of both $\alpha$ and $\beta$ can be described by the so-called dynamo number given as
\begin{equation}
 D =  9 \left(\frac{\Omega H}{v_t}\right)^{2} .
\label{eq11}
\end{equation}
The value of D needs to be larger than a critical value  $D_c $ = 7 for the dynamo to operate as found from numerical simulations of galactic dynamo \citep{Ruzmaikin1988,Arshakian2009}. The critical value $ D_c $ given here depends on the overall thickness of the disk, and this estimate has been determined from thin galactic disks, while for primordial accretion disks we expect $ H/R \sim$ 0.1-0.5. With a dynamo number $D \sim$ 40, an enhancement by a factor of 5 in the $D_c $ can be accommodated. The precise value of $D_c $ should however be checked using numerical simulations. The growth rate for the dynamo is computed as
\begin{equation}
 \Gamma =  D^{1/2} \frac{\beta}{ H^{2}} ,
\label{eq12}
\end{equation}
and the amplification time scale is \citep{Arshakian2009}
\begin{equation}
 t_{growth} = \frac{H}{ \Omega l_t} \sim \frac{1}{ \Omega }  , \label{growth}
\label{eq13}
\end{equation}
implying that $t_{growth}\sim t_{orb}/(2\pi)$, with $t_{orb}$ the orbital timescale of the disk. The final field strength due to the $\alpha-\Omega$ effect is related to  the equipartition field strength $B_*$. The final field strength is then computed via the expression
\begin{equation}
 B_{f} = 0.5 B_*  \sqrt{D/D_c -1} .
\label{eq17}
\end{equation}
The ratio  of $B_f$ / $B_{ini}$ yields the number of e-folding times required to reach the final field strength. 

While the amplification occurs on the timescale $t_{growth}$ defined in Eq.~\ref{growth}, we note that the resulting magnetic field is initially ordered only on scales corresponding to the half-thickness of the disk. Full ordering over the entire disk radius occurs then on the ordering timescale given as \citep{Moss1998}
\begin{equation}
t_{ord} = \frac{R}{\sqrt{ \Gamma \beta}} .
\label{eq18}
\end{equation}
Combining the assumptions above, one can show that $t_{ord}\sim3R/(D^{1/4}v_t$), which is about three times longer than the growth time $t_{growth}$ as the disk is rather thick in our case with  $H/R \sim 0.1- 0.5$  \citep{LatifVolonteri2015}.

\subsection{Jet formation}

The presence of a coherent magnetic field in the disk can lead to the formation of magnetic jets and outflows. Our jet formation model is  based on the framework of \cite{Tan2004}. 
We consider the toroidal component $B_{\phi, s}$ to be the final field strength produced by the $\alpha-\Omega$ dynamo as defined in  Eq.~\ref{eq17}.  $B_{\phi}$ is larger than  the poloidal magnetic field $B_{r}$,  as it is amplified by the shear. The ratio of the poloidal magnetic field $B_{r,s}$ to the toroidal field $B_{\phi,s}$ can be evaluated based on the timescales of buoyancy and convection given as
\begin{equation}
\frac{B_{r,s}}{B_{\phi, s}} \sim  \alpha_{vis}^{1/2}(R) ,
\label{eq12}
\end{equation}
where $\alpha_{vis}$ is the previously introduced disk viscosity parameter.  Requiring that $\vec{\nabla} \cdot \vec{B} =0$, one can  show that 
\begin{equation}
B_{z}/ B_{r} =  \frac{9.0}{16.0} \frac{H}{R} ,
\label{eq13}
\end{equation}
where $B_{z, s}$ is the vertical field inside the disk. Under the assumption of momentum  conservation,  the magnetic Poynting flux is transferred into mechanical energy during the propagation to larger scales. Considering that at a larger distance from the star, the energy of the outflow is dominated by the kinetic energy of the matter, we can express the outflow luminosity as

\begin{equation}
L _{mag}= 0.5 \dot{m}_w v_w^2 ,
\label{outflow}
\end{equation}
where $v_w$ is the terminal velocity of the outflow and $\dot{m}_w$ is the mass outflow rate. According to theoretical models for the magnetically driven outflows, $v_w$ can be  estimated from the escape velocity $v_{esc}$ as \citep{Shu2000,Konigl2000} 
\begin{equation}
v_{esc}= \sqrt{2GM/R} .
\label{eq14}
\end{equation}

In the following, we will adopt a differential formulation by considering the vertical Poynting flux through a disk annulus with radius $R$ and thickness $dR$. Its contribution to the jet luminosity is thus given as \citep{Lovelace2002}
\begin{equation}
\frac{dL_{mag}}{dR} = B_{\phi, s} (R)B_{z, s} (R)  \Omega R^2 
\label{dlum}
\end{equation}
for quasi-static conditions. The differential magnetic luminosity of the annulus $dL_{mag}$ drives a differential outflow rate $d\dot{m}_w$, which is related via\begin{equation}
dL_{mag}=\frac{1}{2}d\dot{m}_w v_w^2 ,
\end{equation}
where the terminal velocity $v_w$ is given as the escape velocity at radius $R$. Integrating the expressions for $dL_{mag}$ and $d\dot{m}_w$ yields the magnetic luminosity and mass outflow rate within radius $R$. In particular, we define
\begin{equation}
L_{mag}(R)=\int_0^R dL_{mag}(R)
\end{equation}
and
\begin{equation}
\dot{m}(R)=\int_0^R d\dot{m}(R) .
\end{equation}

\section{Results}\label{results}
In the following, we employ the model outlined above to explore the disk properties and their impact on magnetic field amplification and jet formation for two characteristic cases, a Pop. III star and a supermassive star. In the first case, we consider a typical mode of Pop.~III star formation, with cooling driven by molecular hydrogen and moderate accretion rates of $\sim10^{-3}$~M$_\odot$~yr$^{-1}$ \citep[e.g.][]{Abel2002,Bromm2002, Yoshida08}. As a typical mass, we adopt $\sim10$~M$_{\odot}$ during the protostellar accretion phase \citep[see e.g.][]{LatifPopIII13}, which may increase to $>100$~M$_{\odot}$ until the star reaches the main sequence. 

We compare this scenario with the accretion disk surrounding a supermassive protostar of $\sim10^5$~M$_\odot$, which is expected to form in more massive primordial halos in the presence of strong UV backgrounds \citep{Hosokawa2013,Latif2013d,Schleicher13,Regan2014,Latif2014ApJ, Ferrara14}. In this regime, simulations have derived characteristic accretion rates of $10^{-1}$~M$_\odot$~yr$^{-1}$. Here, we aim to quantify the characteristic magnetic field strength, jet luminosities and mass outflow rates in both cases. The latter can be particularly valuable to determine the strength of protostellar feedback from such protostars.


\subsection{Pop. III star}
\begin{figure*}
\hspace{-6.0cm} 
\centering
\begin{tabular}{c c}
\begin{minipage}{6cm}
\vspace{0.2cm}
\includegraphics[scale=0.7]{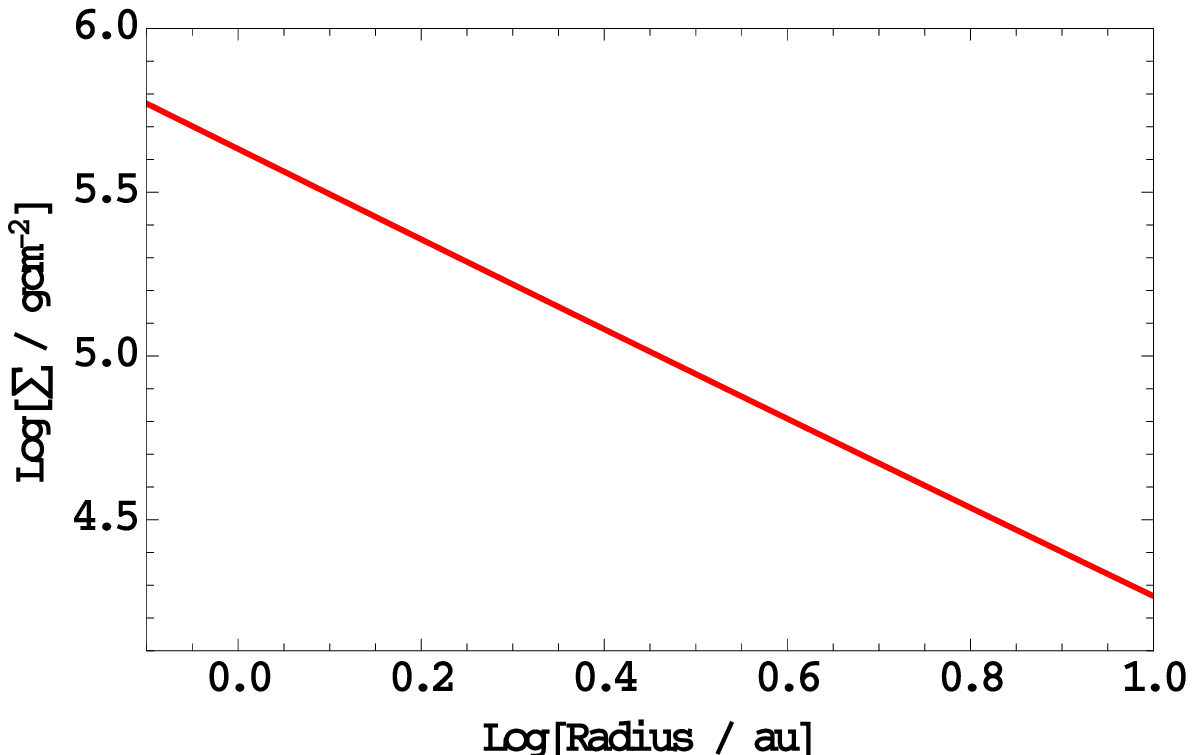}
\end{minipage} &
\begin{minipage}{6cm}
\hspace{2.4cm}
\includegraphics[scale=0.7]{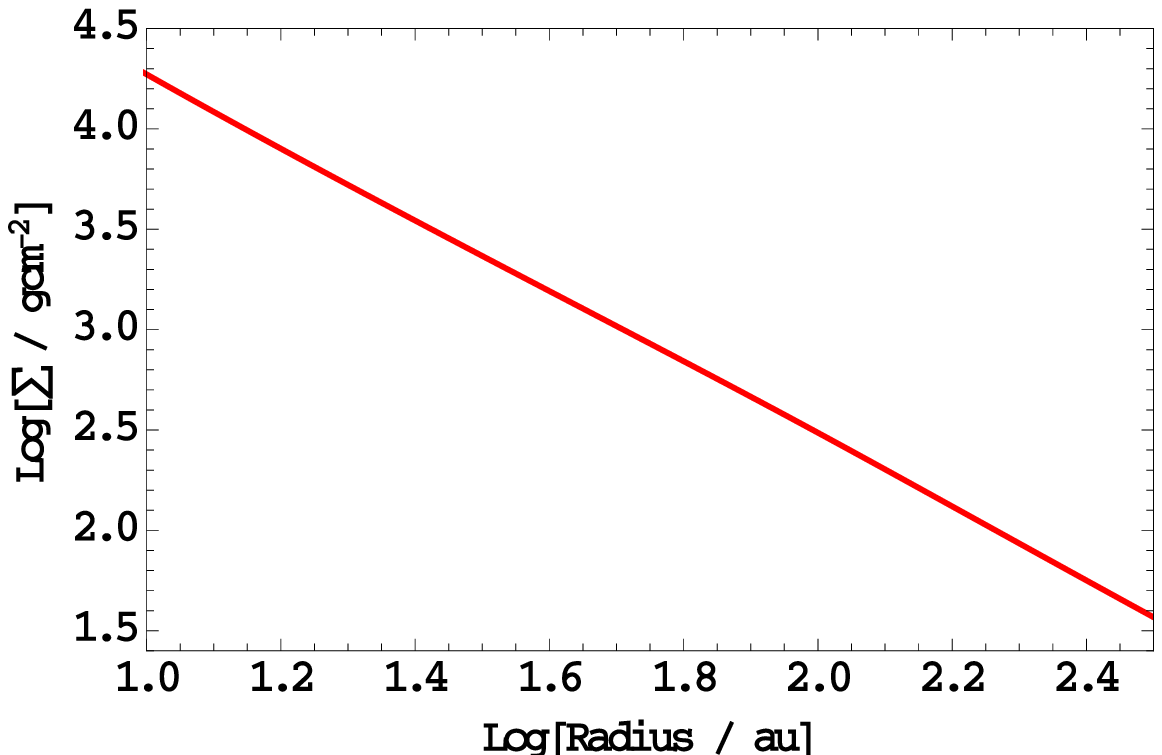}
\end{minipage} \\
\begin{minipage}{6cm}
\vspace{0.2cm}
\includegraphics[scale=0.7]{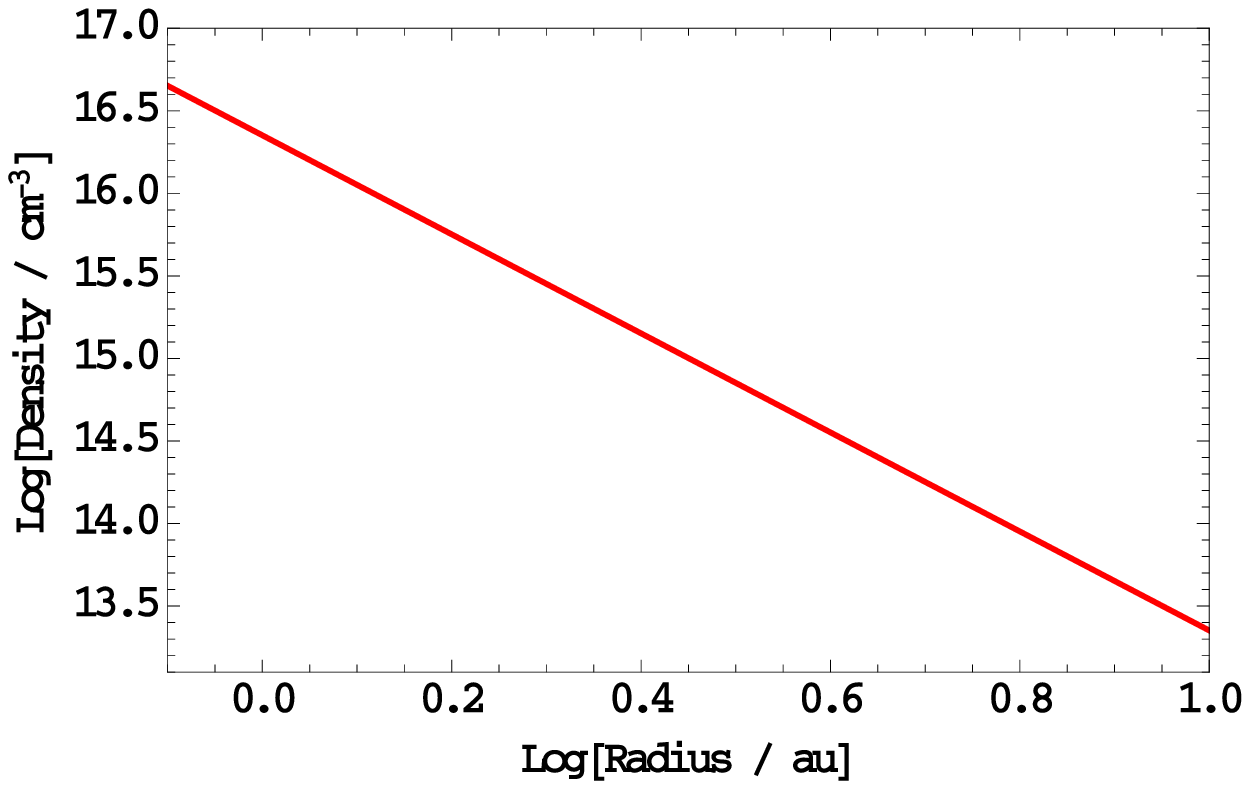}
\end{minipage} &
\begin{minipage}{6cm}
\hspace{2.3cm}
\includegraphics[scale=0.7]{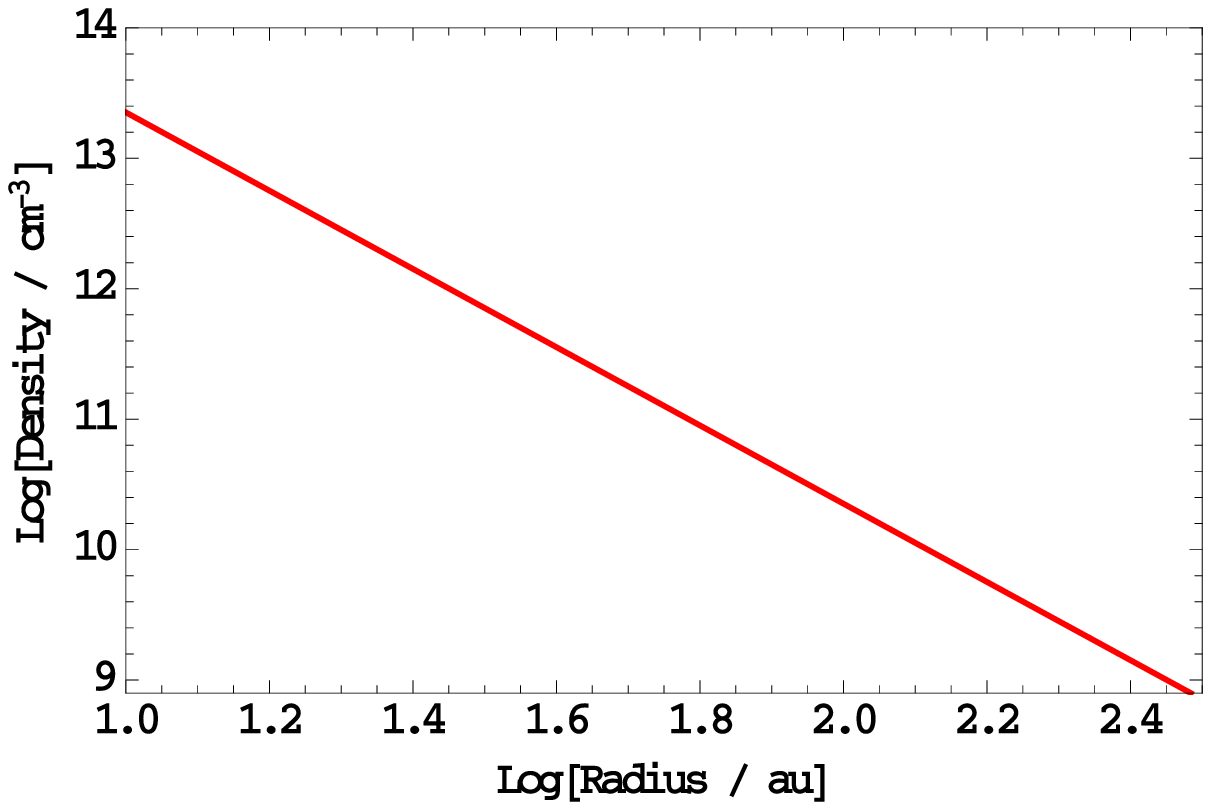}
\end{minipage} \\
\begin{minipage}{6cm}
\vspace{-0.2cm}
\includegraphics[scale=0.7]{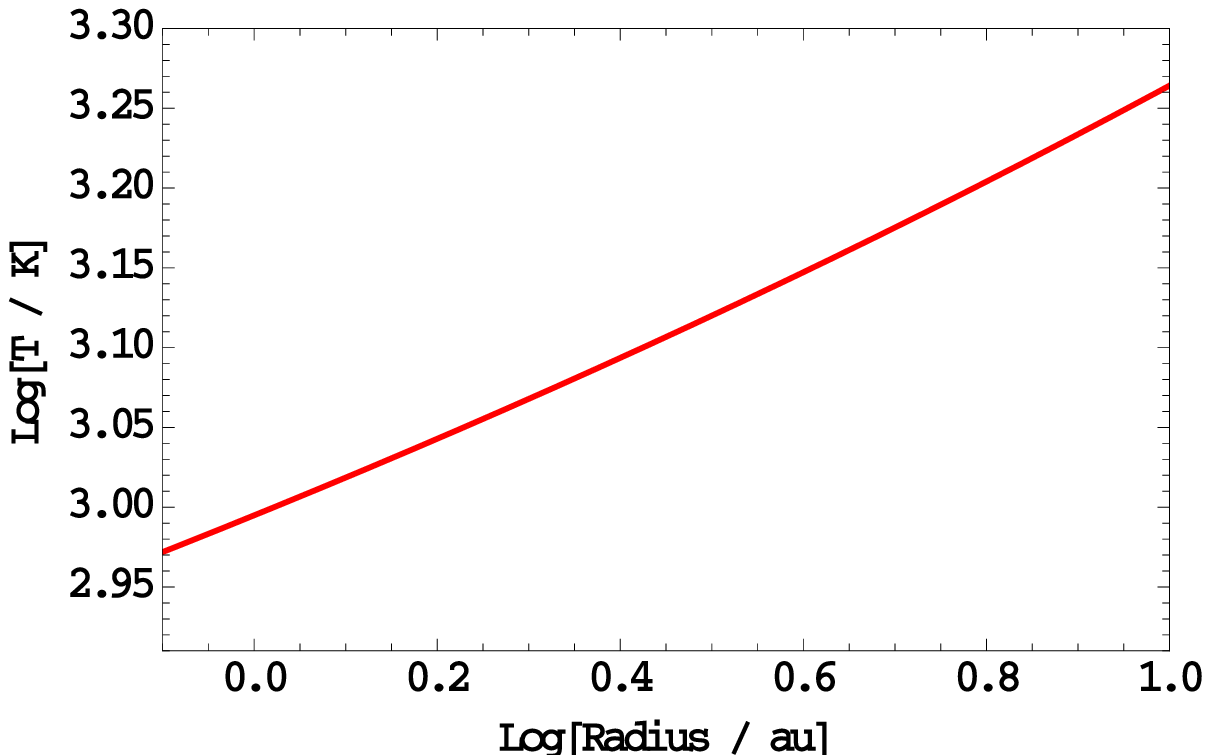}
\end{minipage} &
\begin{minipage}{6cm}
\hspace{2.3cm}
\includegraphics[scale=0.7]{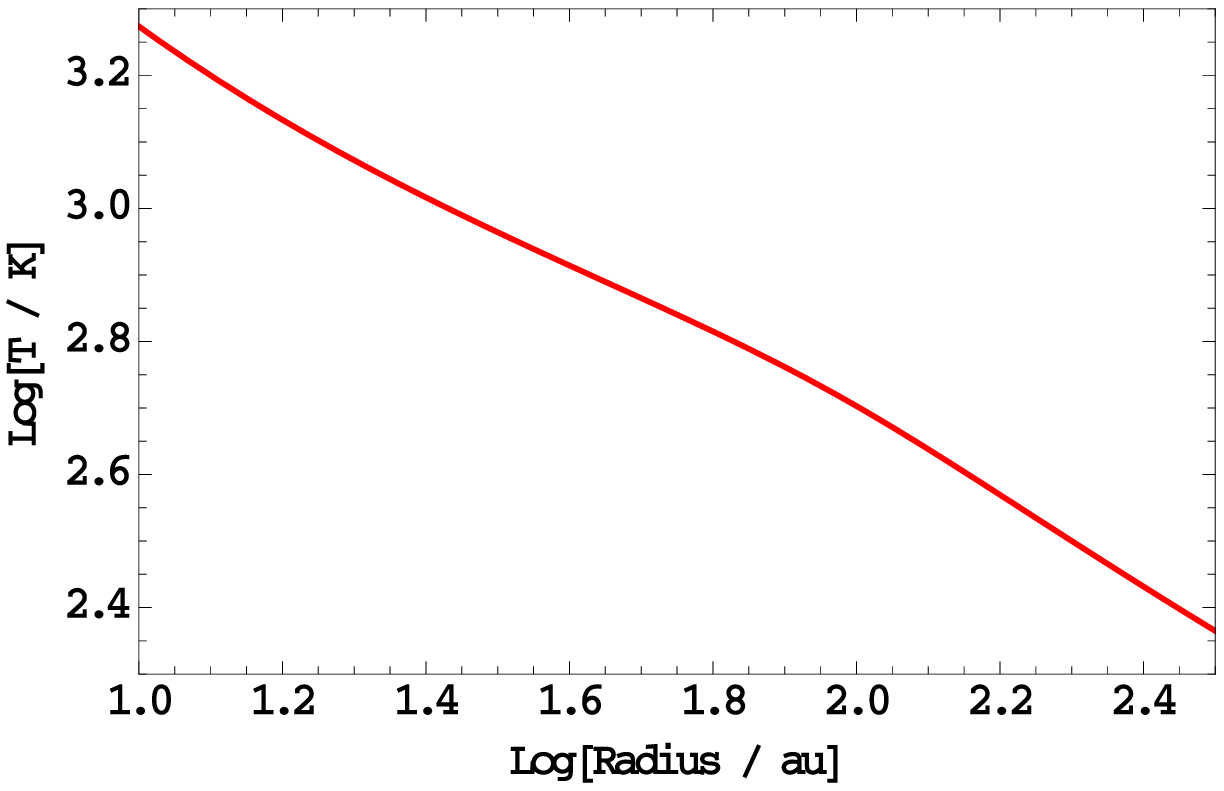}
\end{minipage}
\end{tabular}
\caption{The surface density (top panel), gas midplane density (mid panel) and the gas temperature  (bottom panel) of the disk for a $\rm 10~M_{\odot}$ central star with an accretion rate of $10^{-3}$~M$_\odot$~yr$^{-1}$. The left panels represent the CIE cooling regime, while the right panels show the regime of $\rm H_2$ line cooling.} 
\label{fig1}
\end{figure*}

In the case of a  Pop. III star with moderate accretion rates, the cooling on larger scales is due to molecular hydrogen line cooling, while at smaller scales and for densities above $\sim10^{14}$~cm$^{-3}$, the cooling is driven via collisionally-induced emission (CIE) of molecular hydrogen. In the following, we describe the disk structure obtained in both regimes, employing the cooling rates as described by \citet{Latif2014Disk}. For the protostar, we adopt a generic mass of $10$~M$_\odot$ with an accretion rate of $\sim10^{-3}$~M$_\odot$~yr$^{-1}$, as found in numerical simulations by \citet{LatifPopIII13}. The disk structure obtained with our model is given in Fig.~\ref{fig1}, where we show the density and temperature structure both in the molecular cooling and the CIE cooling regime. 

The surface density of the disk increases from $\sim30$~g~cm$^{-2}$ at $300$~AU to $\sim5\times10^5$~g~cm$^{-2}$ at $1$~AU, and follows  $\Sigma \propto R^{-1.5}$.  The mid-plane density of the disk at $300$~AU is $10^{10}$~cm$^{-3}$, which increases to $\sim10^{17}$~cm$^{-3}$ at $1$~AU. The temperature increases in the H$_2$ cooling regime towards smaller scales until it reaches $\sim2000$~K, where CIE cooling sets in on scales of $\sim10$~AU. From that point, the temperature remains approximately constant, with a minor decrease towards $1$~AU.

\begin{figure*}
\hspace{-6.0cm}
\centering
\begin{tabular}{c c}
\begin{minipage}{6cm}
\vspace{-0.2cm}
\includegraphics[scale=0.7]{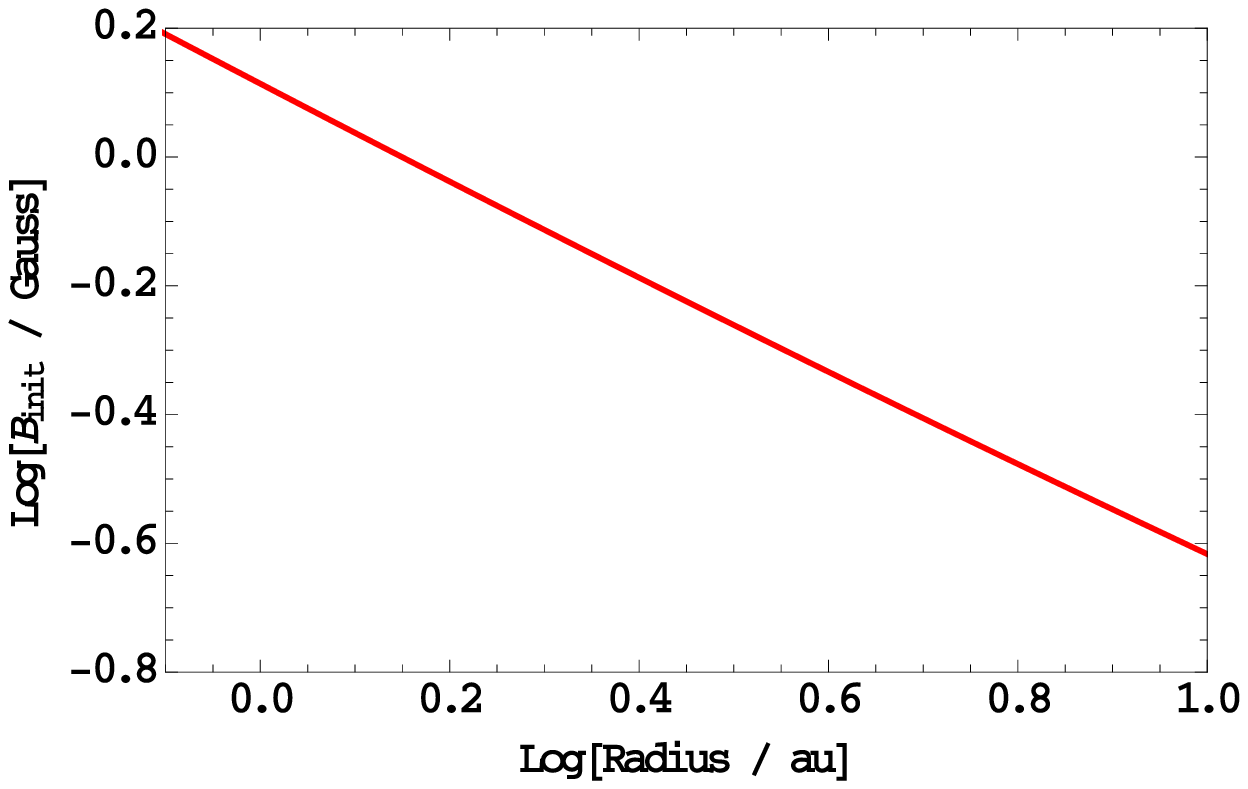}
\end{minipage} &
\begin{minipage}{6cm}
\hspace{2.3cm}
\includegraphics[scale=0.7]{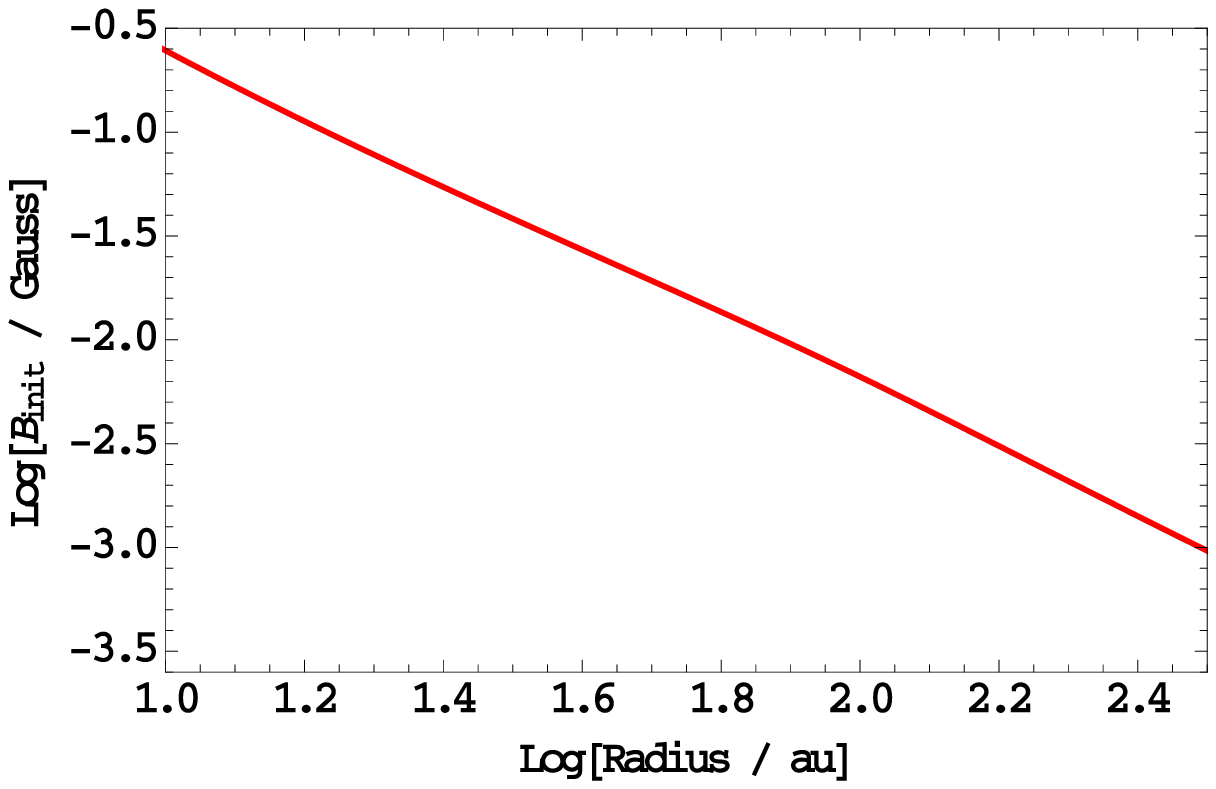}
\end{minipage} \\
\begin{minipage}{6cm}
\vspace{0.2cm}
\includegraphics[scale=0.7]{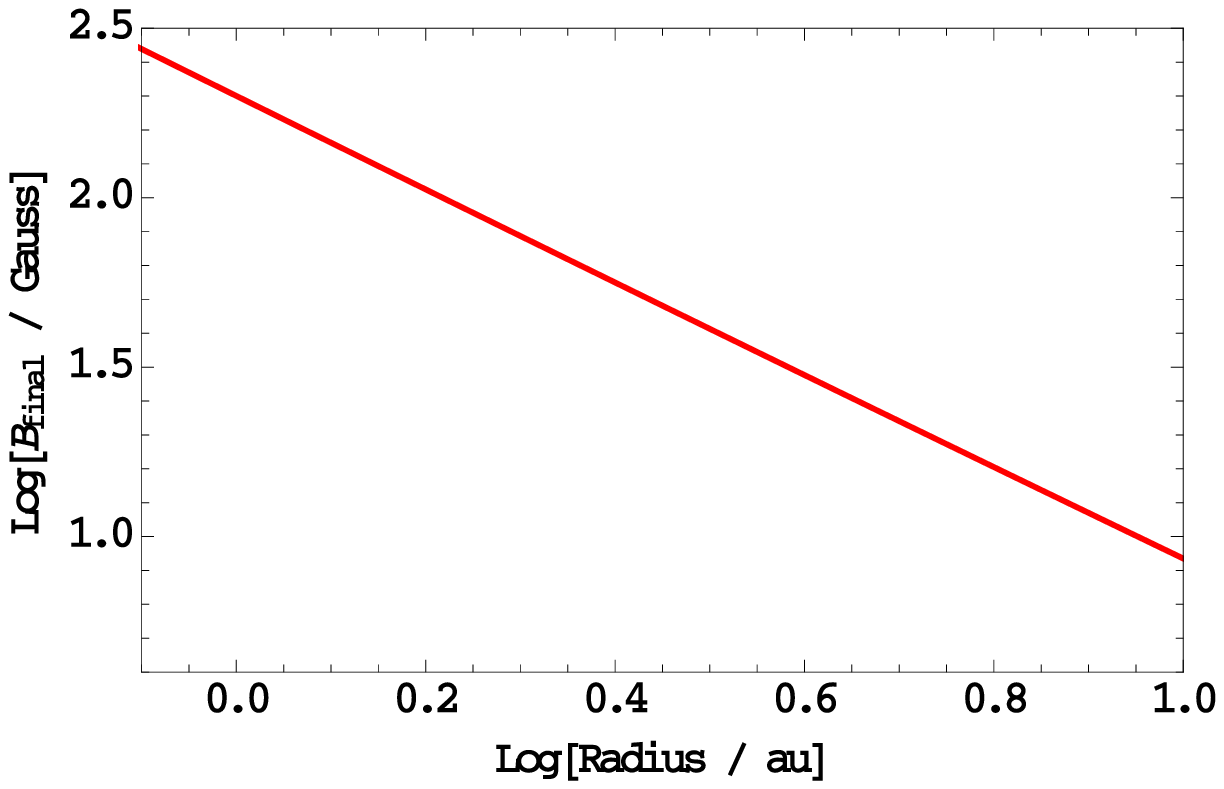}
\end{minipage} &
\begin{minipage}{6cm}
\hspace{2.3cm}
\includegraphics[scale=0.7]{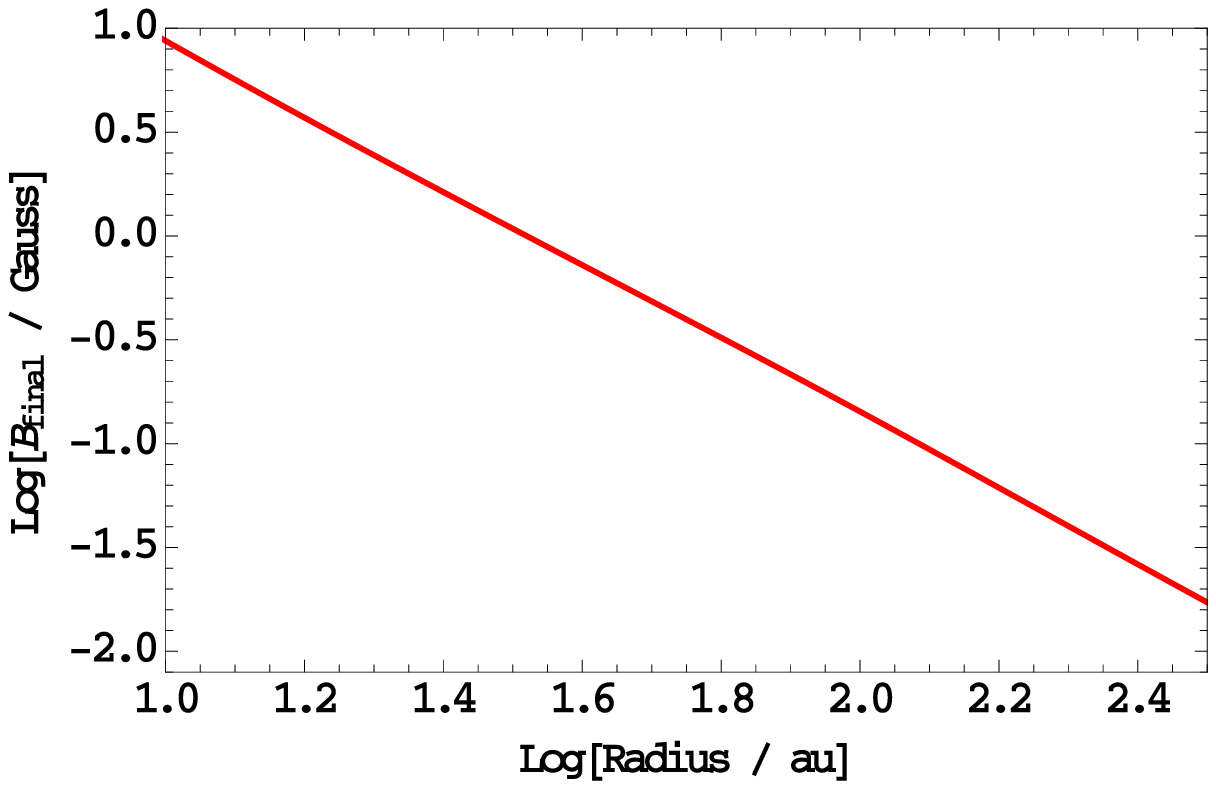}
\end{minipage}\\
\begin{minipage}{6cm}
\vspace{0.2cm}
\includegraphics[scale=0.7]{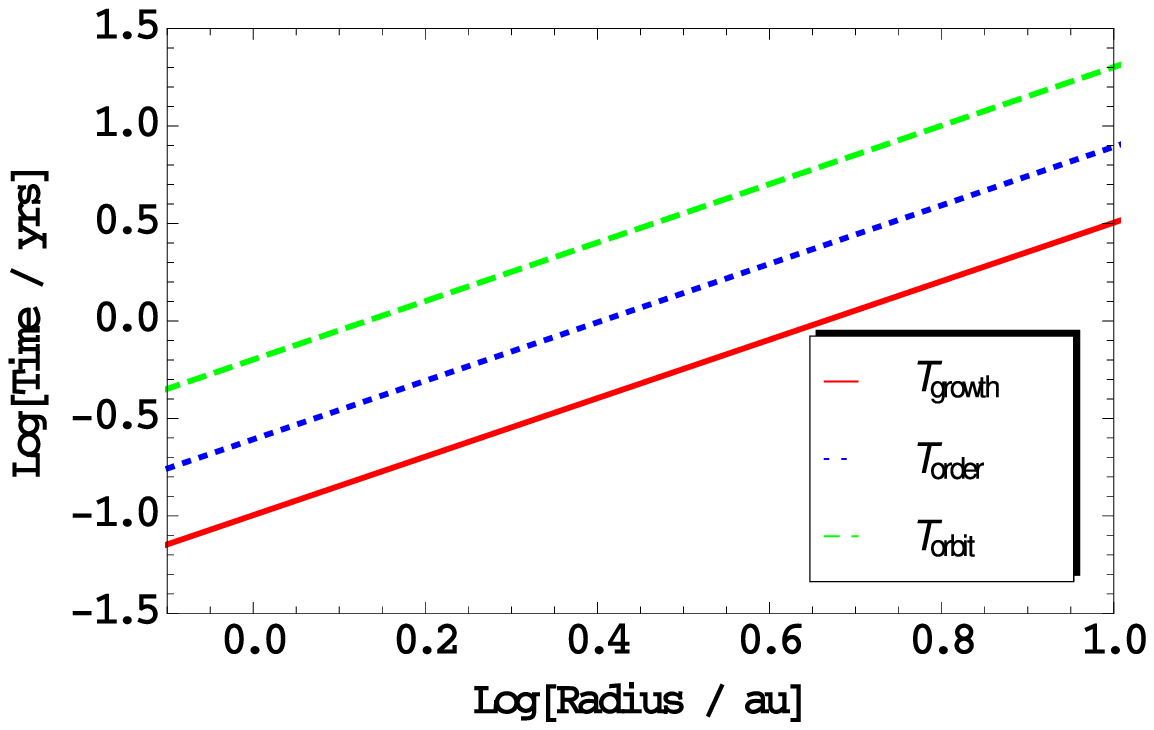}
\end{minipage} &
\begin{minipage}{6cm}
\hspace{2.3cm}
\includegraphics[scale=0.7]{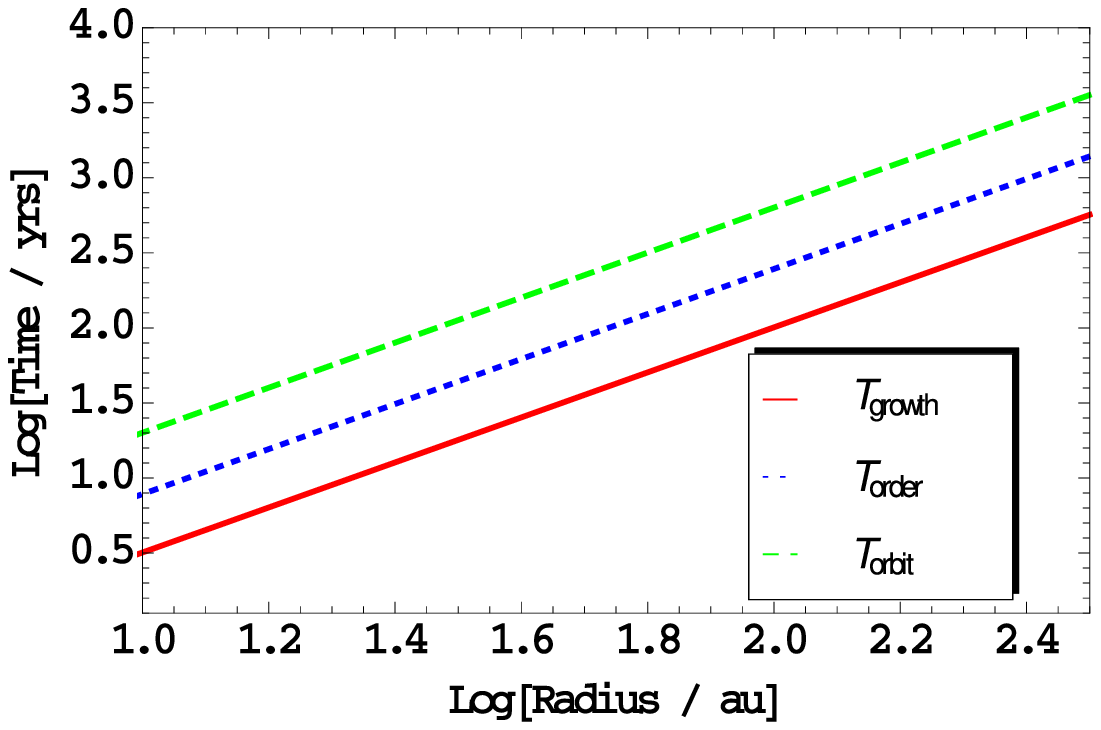}
\end{minipage} 
\end{tabular}
\caption{The initial magnetic field strength $B_{init}$ generated by the small-scale dynamo (the top panel) and the final magnetic field strength $B_f$ generated by the $\alpha- \Omega$ dynamo (mid panel) for a $\rm 10~M_{\odot}$ central star  with an accretion rate of $10^{-3}$~M$_\odot$~yr$^{-1}$. On the bottom panel, we show the orbital timescale, the growth time of the $\alpha-\Omega$ dynamo and the ordering timescale. The left panels represent the CIE cooling regime, while the right panels show the regime of $\rm H_2$ line cooling.}
\label{fig2}
\end{figure*}

The magnetic field strength obtained in such a disk via the dynamo process is shown in Fig.~\ref{fig2}, including the coherent magnetic field component produced by the small-scale dynamo ($B_{init}$), as well as the final field strength obtained via the $\alpha-\Omega$ dynamo.  From the small-scale dynamo, we obtain a characteristic field strength $B_{init}$ of $10^{-3}-10^{0.2}$~G on scales ranging from $300$~AU down to $1$~AU. The latter is due to the high gas densities of at least $10^9$~cm$^{-3}$ as well as the high temperatures in the gas, reflected by an amount of turbulent energy which is  about 50\% of the thermal energy, and consistent with estimates from numerical simulations \citep{Greif12,LatifPopIII13}. The magnetic field strength is further enhanced by the $\alpha-\Omega$ dynamo, with a final field strength $B_f$ of $10^{-1.8}-10^{+2.4}$~G.

We further show the orbital timescale, the growth time and the ordering timescale within the same figure. As mentioned in section~\ref{magnetic}, the growth time corresponds to about $1/(2\pi)$ of the orbital time. Depending on the scale, the ratio between the initial and final magnetic field strength varies between $\sim50$ on scales of $1$~AU and $\sim5$ on scales of $300$~AU, corresponding to $2.7-5$ e-folding times until saturation, and the ordering timescale may add a few more e-folds. Considering the characteristic timescale for accretion, $t_{acc}=M_*/\dot{M}\sim10^4$~yrs, we find that the timescale to reach saturation becomes comparable to the accretion time on scales of $\sim300$~AU and is considerably shorter on smaller scales. As a result, our assumption of saturation is justified at least until scales of $100$~AU.

\begin{figure*}
\hspace{-6.0cm}
\centering
\begin{tabular}{c c}\begin{minipage}{6cm}
\vspace{0.2cm}
\includegraphics[scale=0.7]{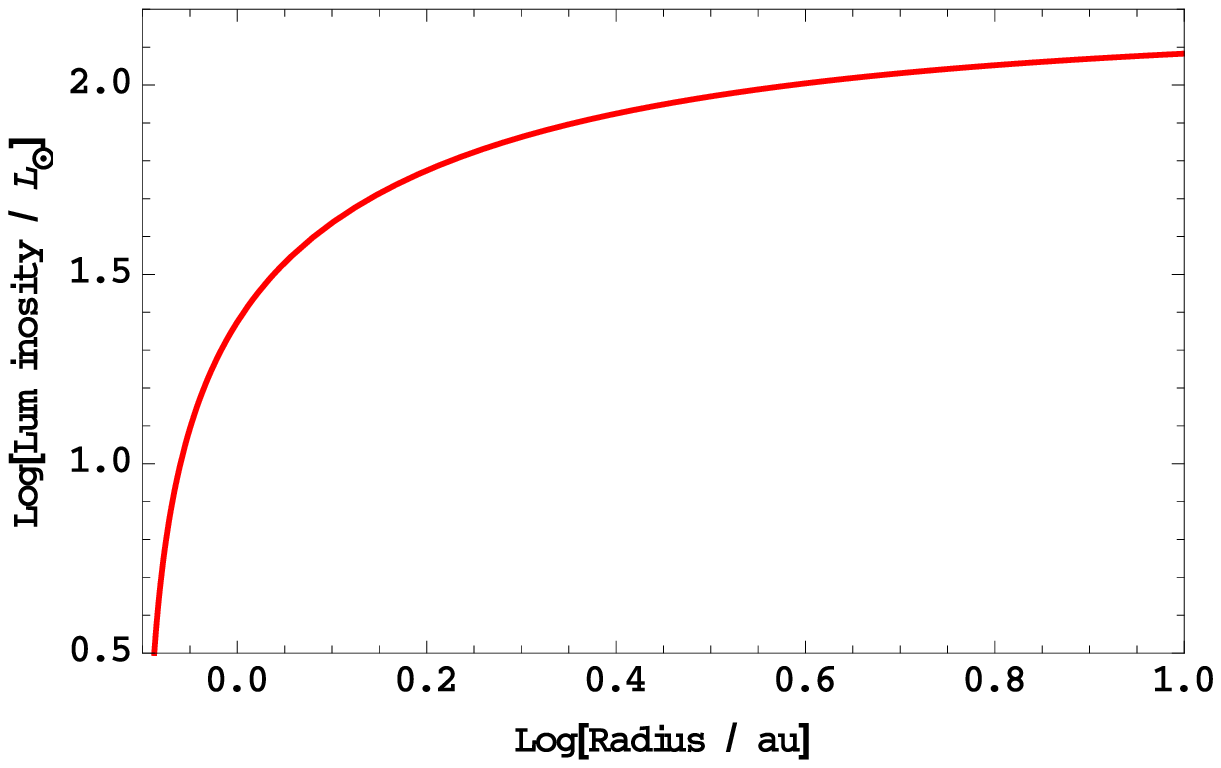}
\end{minipage} &
\begin{minipage}{6cm}
\hspace{2.3cm}
\includegraphics[scale=0.7]{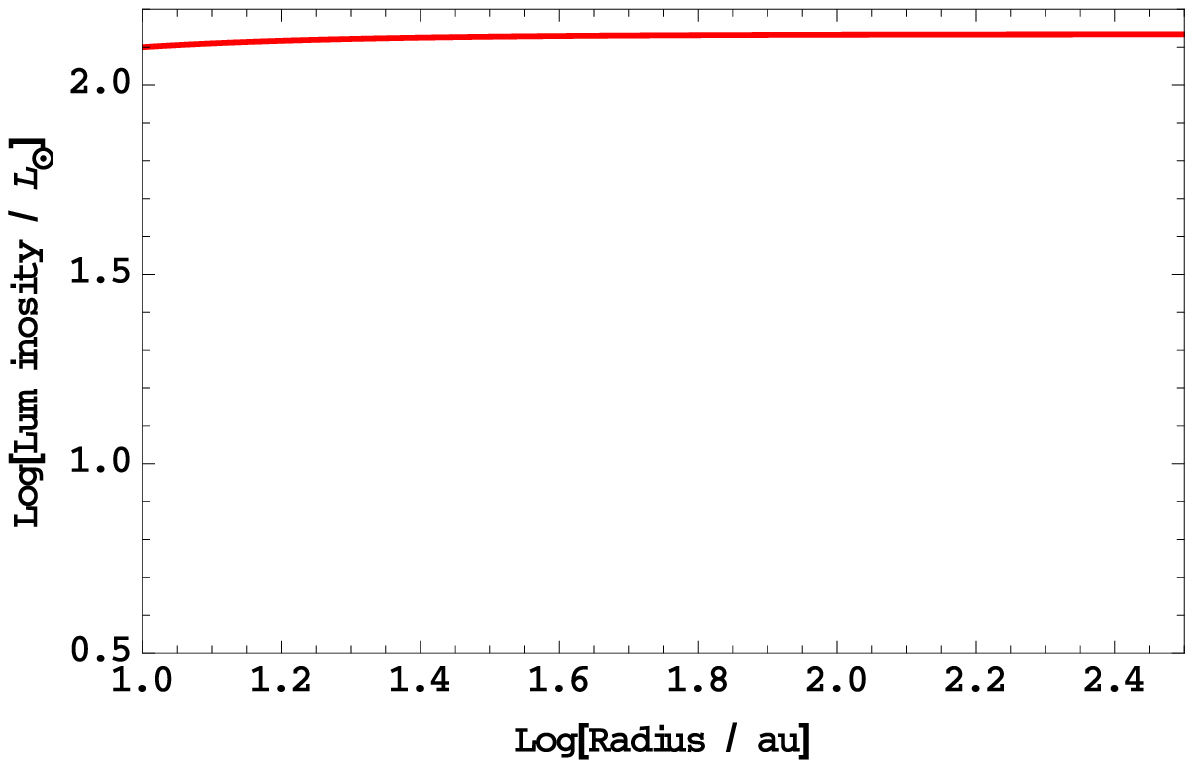}
\end{minipage} \\
\begin{minipage}{6cm}
\vspace{0.2cm}
\includegraphics[scale=0.7]{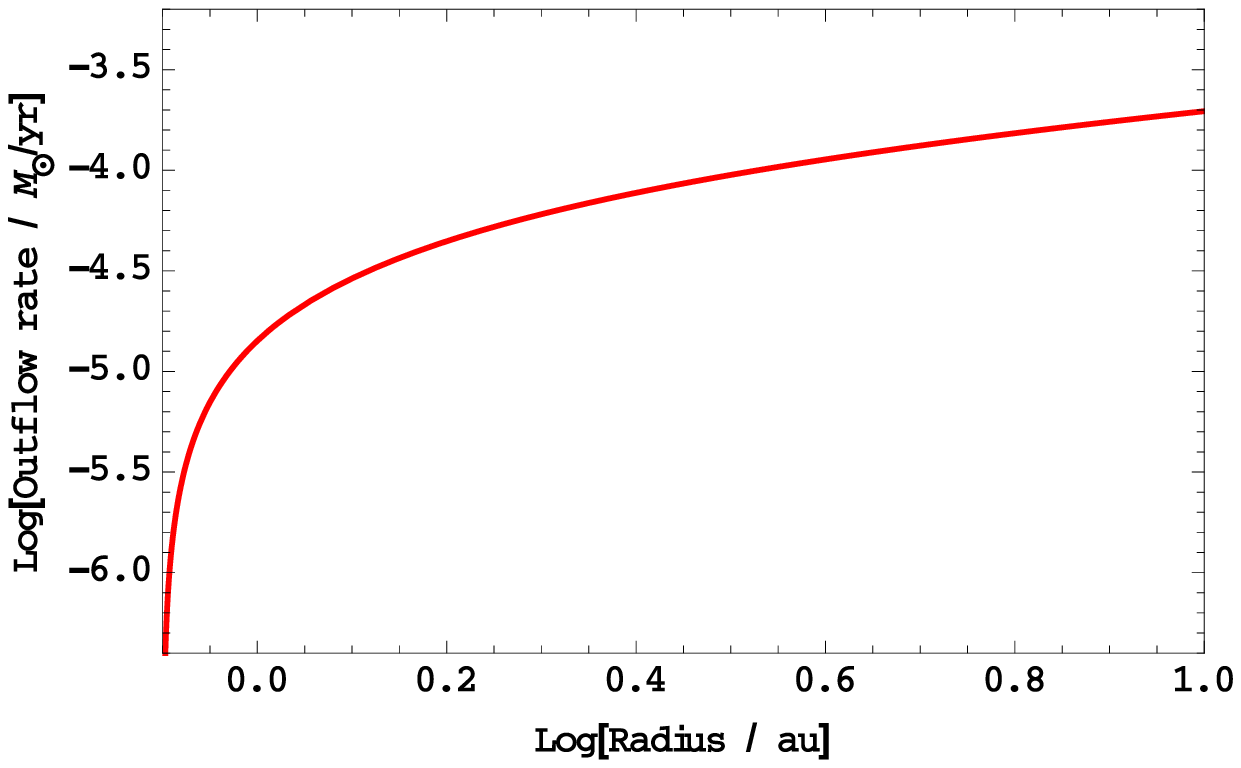}
\end{minipage} &
\begin{minipage}{6cm}
\hspace{2.3cm}
\includegraphics[scale=0.7]{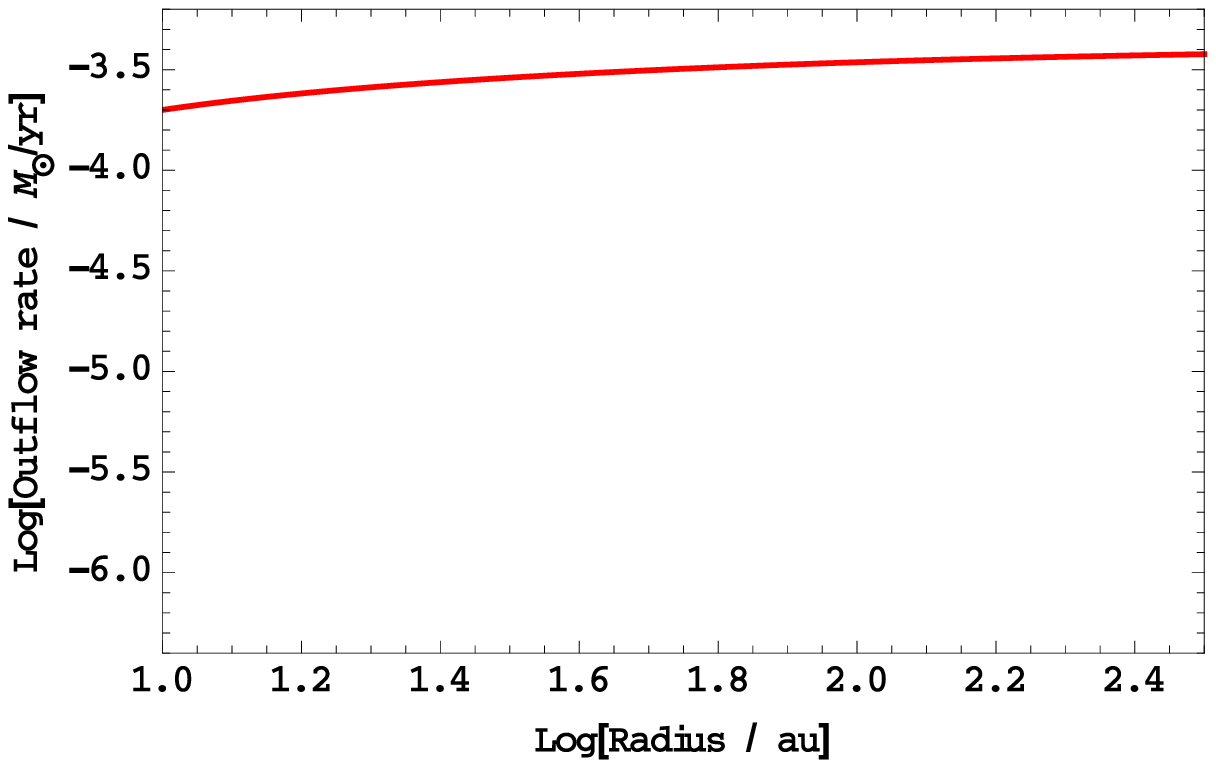}
\end{minipage}
\end{tabular}
\caption{Expected properties of the magnetic outflow for a $10$~M$_\odot$ star  with an accretion rate of $10^{-3}$~M$_\odot$~yr$^{-1}$. The magnetic luminosity  is depicted in the top panel while the bottom panel shows the mass outflow rate within a given radius $R$. The left panels represent the CIE cooling regime, while the right panels show the regime of $\rm H_2$ line cooling. }
\label{fig3}
\end{figure*}

The resulting properties of the magnetic outflow are given in Fig.~\ref{fig3}, showing the magnetic luminosity and the mass outflow rate within a given radius $R$ both in the CIE cooling and the molecular cooling regime. It is clearly noticeable that both quantities are dominated by contributions from the inner $10$~AU where the disk structure is determined by CIE cooling. The magnetic power to drive the outflow reaches up to $100$~L$_\odot$ at $10$~AU, corresponding to a mass outflow of $10^{-3.7}$~M$_\odot$~yr$^{-1}$. While the latter corresponds to a minor reduction of the mass reaching the central object, it provides a relevant feedback mechanism towards larger scales with a characteristic velocity of $\sim 60$~km/s, which can  drive the magnetization on larger scales with a more coherent magnetic field.

\subsection{Supermassive star} 
Now, we consider the accretion disk surrounding a supermassive star with $10^5$~M$_\odot$ and an accretion rate of $0.1$~M$_\odot$~yr$^{-1}$, as motivated via numerical simulations \citep{Latif2013d,Regan2014,Latif2014ApJ}. The thermal properties of the disk structure were previously explored by \citet{Latif2015c}, finding that the disk is dominated by molecular cooling on large radii, while a transition to the atomic cooling regime may occur on scales of $\sim300$~AU as a result of viscous heating and the collisional dissociation of atomic hydrogen.

\begin{figure*}
\hspace{-6.0cm}
\centering
\begin{tabular}{c c}
\begin{minipage}{6cm}
\vspace{0.2cm}
\includegraphics[scale=0.7]{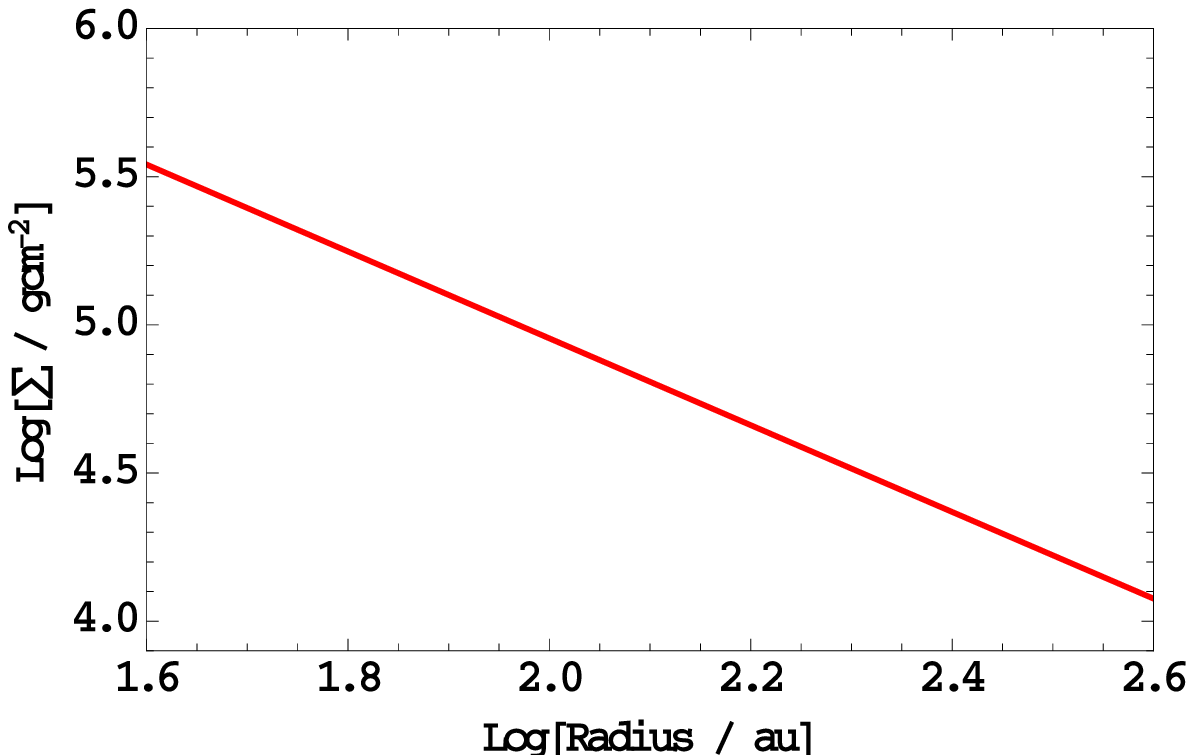}
\end{minipage} &
\begin{minipage}{6cm}
\hspace{2.3cm}
\includegraphics[scale=0.7]{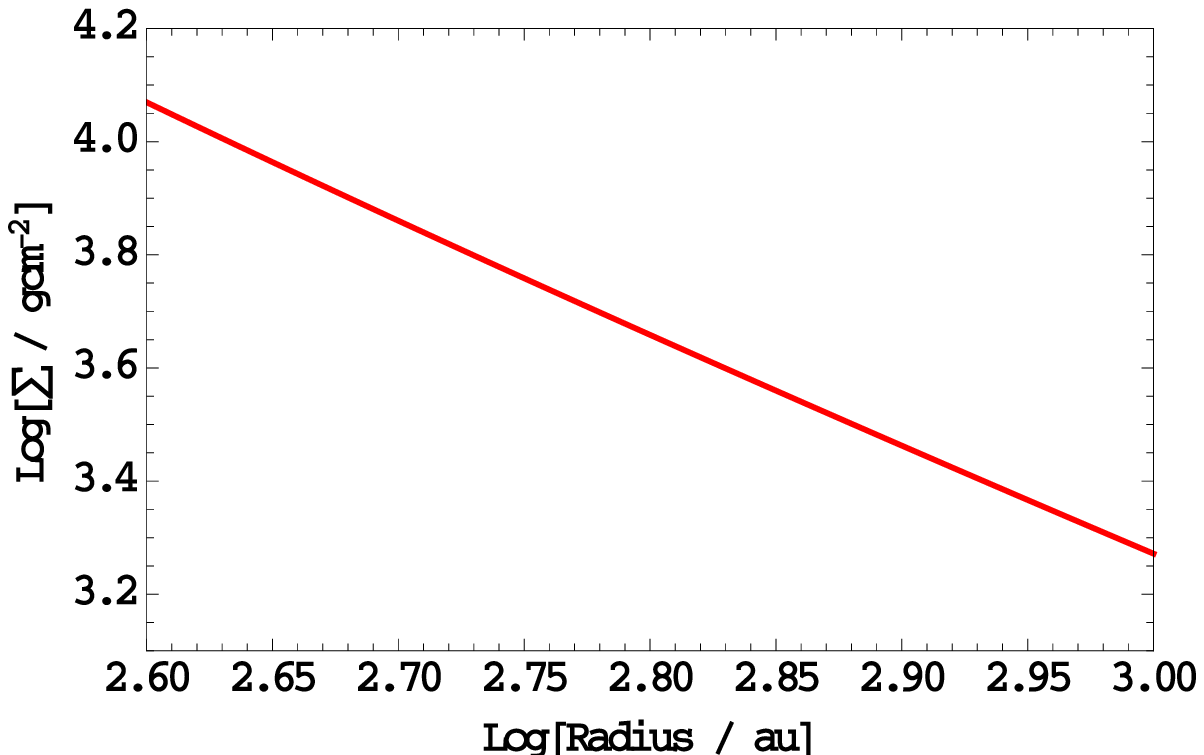}
\end{minipage} \\
\begin{minipage}{6cm}
\vspace{0.2cm}
\includegraphics[scale=0.7]{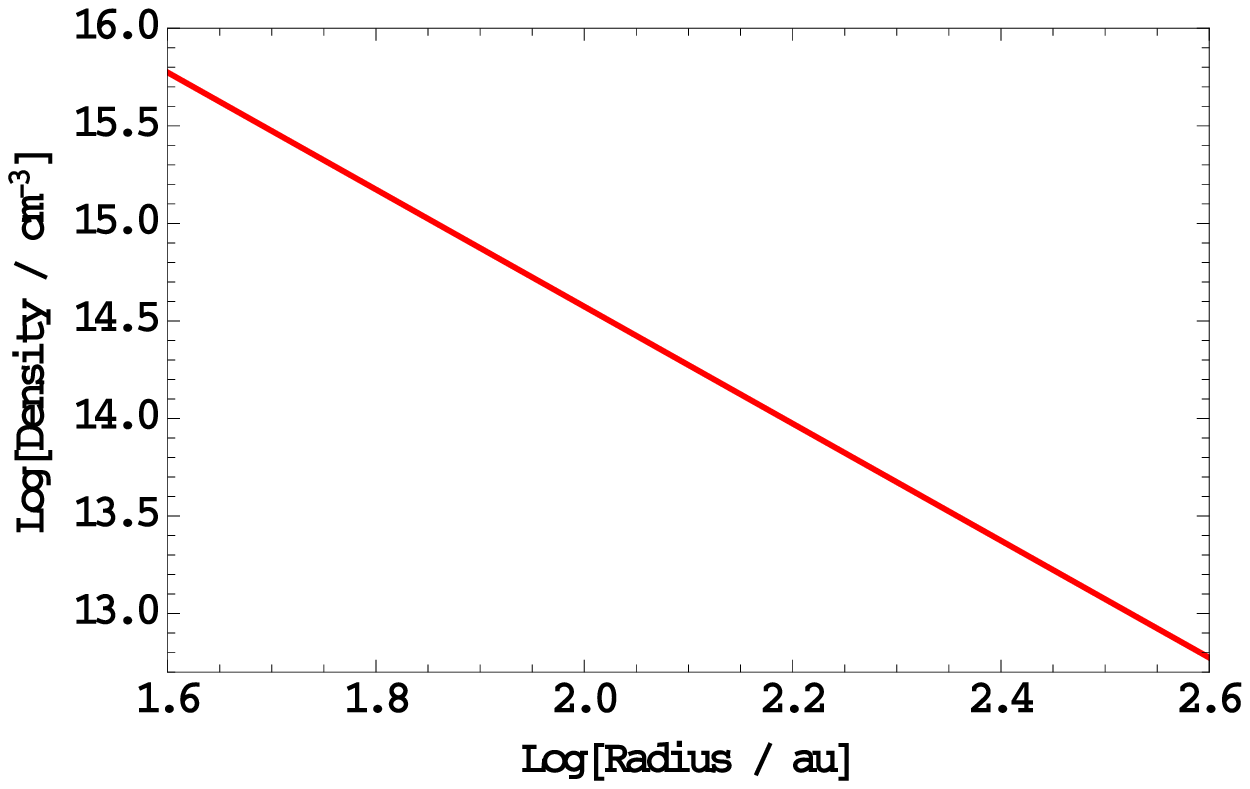}
\end{minipage} &
\begin{minipage}{6cm}
\hspace{2.3cm}
\includegraphics[scale=0.7]{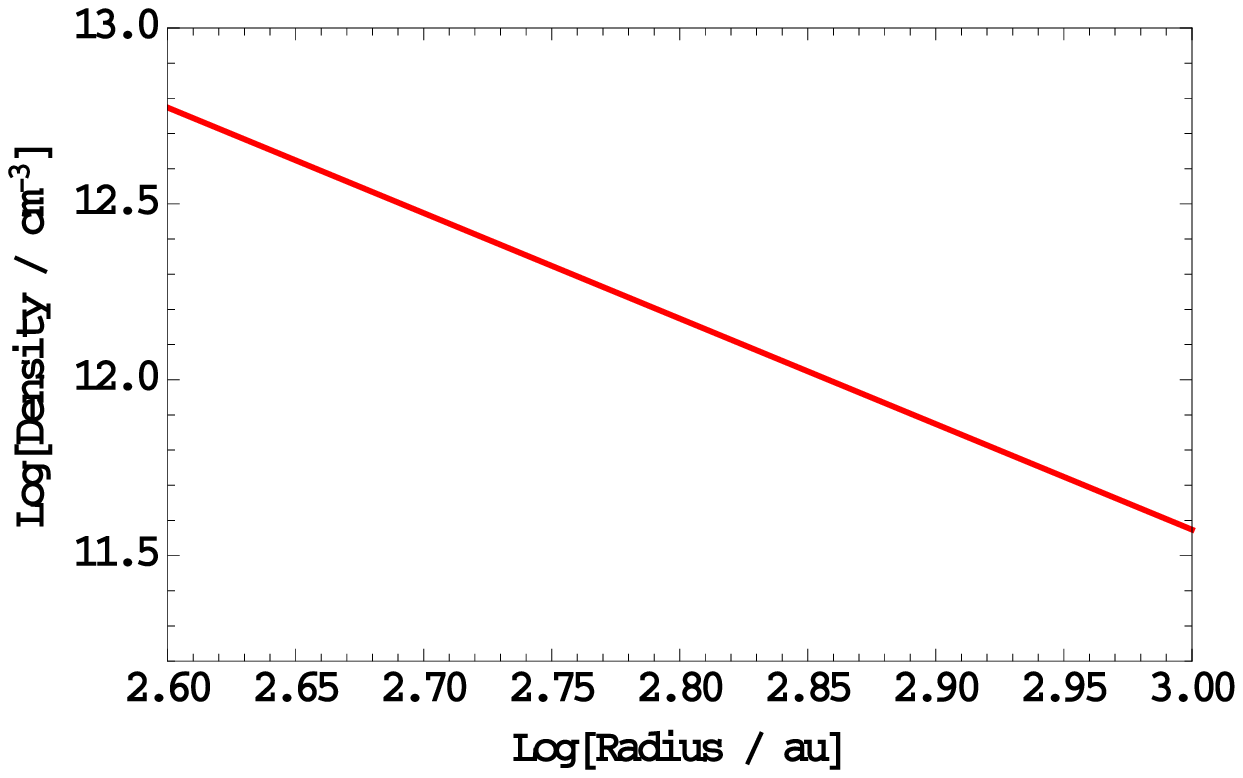}
\end{minipage} \\
\begin{minipage}{6cm}
\vspace{-0.2cm}
\includegraphics[scale=0.7]{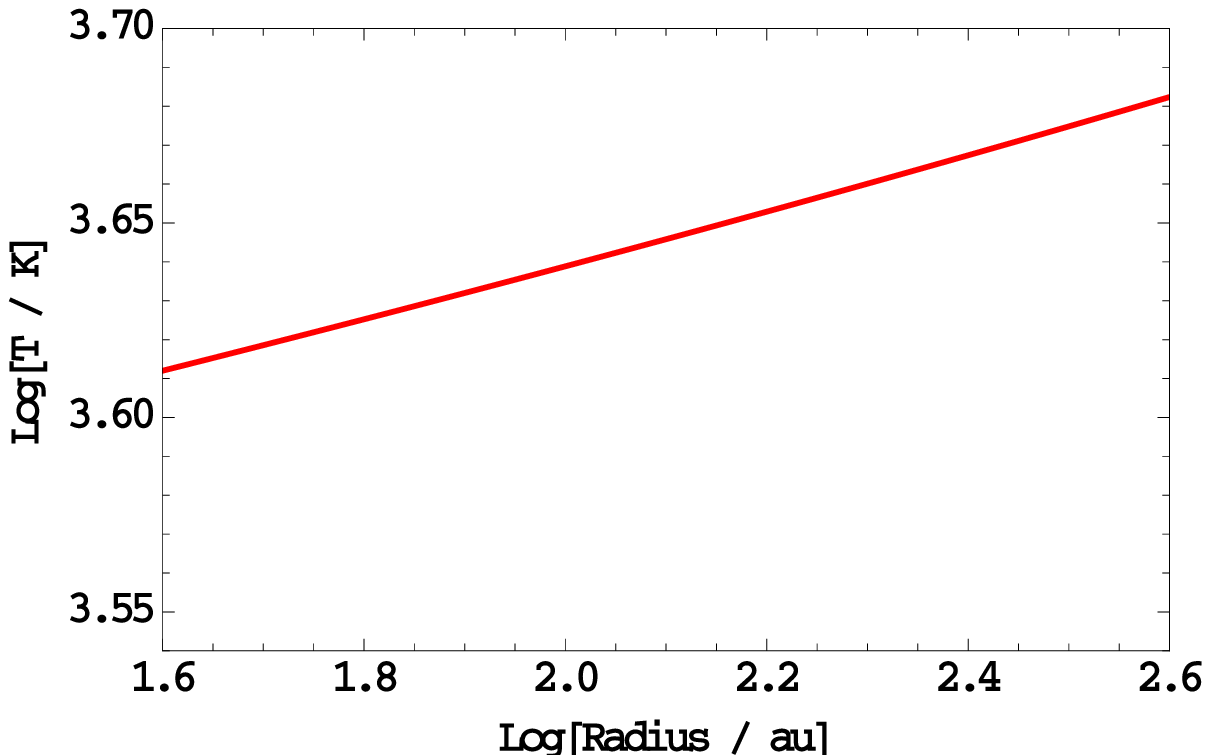}
\end{minipage} &
\begin{minipage}{6cm}
\hspace{2.3cm}
\includegraphics[scale=0.7]{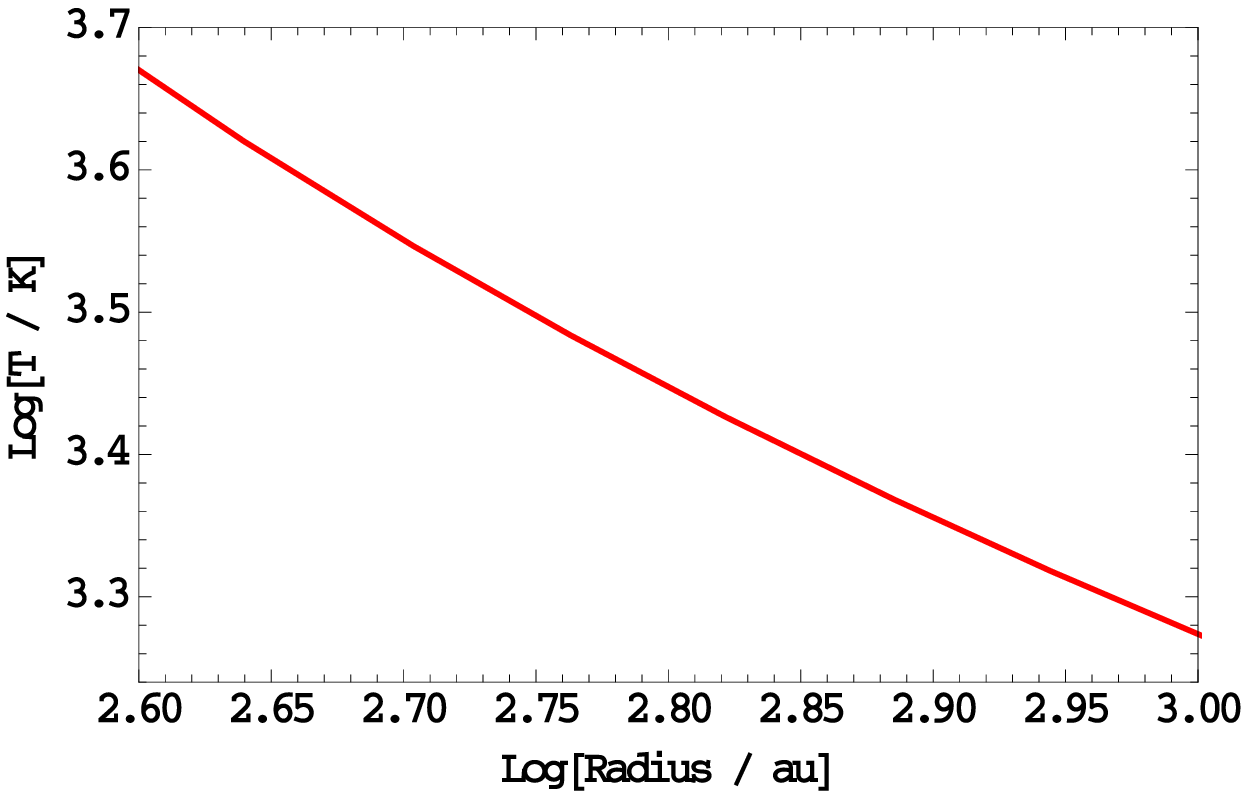}
\end{minipage}
\end{tabular}
\caption{The surface density (top panel), gas midplane density (mid panel)  and the gas temperature (bottom panel) of the disk  are shown for $\rm 10^5~M_{\odot}$ central star  with an accretion rate of $10^{-1}$~M$_\odot$~yr$^{-1}$. The left panels show the atomic cooling regime in the inner parts of the disk, while the right panels show the $\rm H_2$ cooling regime on larger scales.}
\label{fig4}
\end{figure*}

The resulting disk structure in terms of density and temperature is plotted in Fig.~\ref{fig4}. Due to the high accretion rates, the disk is considerably denser in this case, with the surface density ranging from $10^{3.2}$~g~cm$^{-2}$ at $1000$~AU to $10^{5.5}$~g~cm$^{-2}$ at $30$~AU. Similarly, the gas mid-plane density  ranges from $\sim10^{12}$~cm$^{-3}$ on large radii to $10^{16}$~cm$^{-3}$ in the interior parts of the disk. In the molecular cooling regime, the temperature of the disk increases towards smaller scales as a result of viscous heating, leading to temperatures of $\sim4000$~K where molecular hydrogen is efficiently dissociated. Within the central $300$~AU, the gas temperature is almost constant, with a very minor decrease towards higher densities.

\begin{figure*}
\hspace{-6.0cm}
\centering
\begin{tabular}{c c}
\begin{minipage}{6cm}
\vspace{-0.2cm}
\includegraphics[scale=0.7]{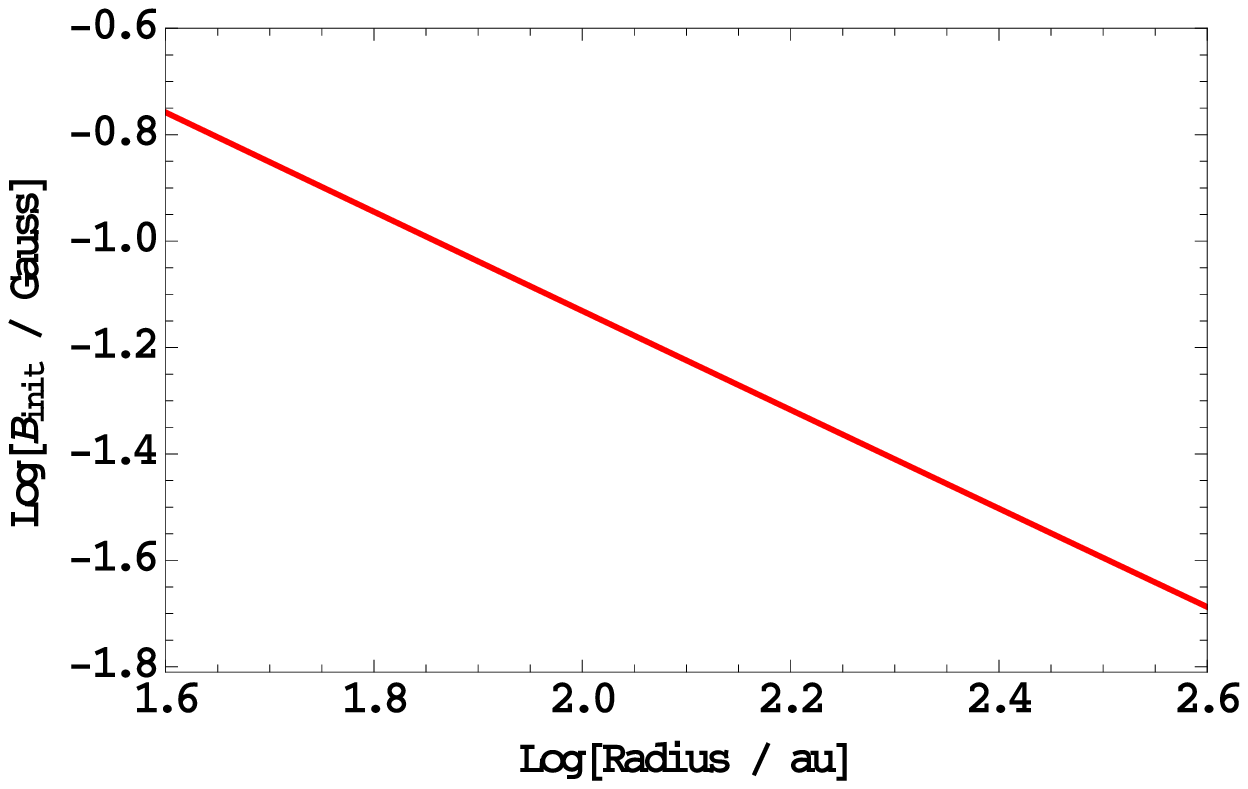}
\end{minipage} &
\begin{minipage}{6cm}
\hspace{2.3cm}
\includegraphics[scale=0.7]{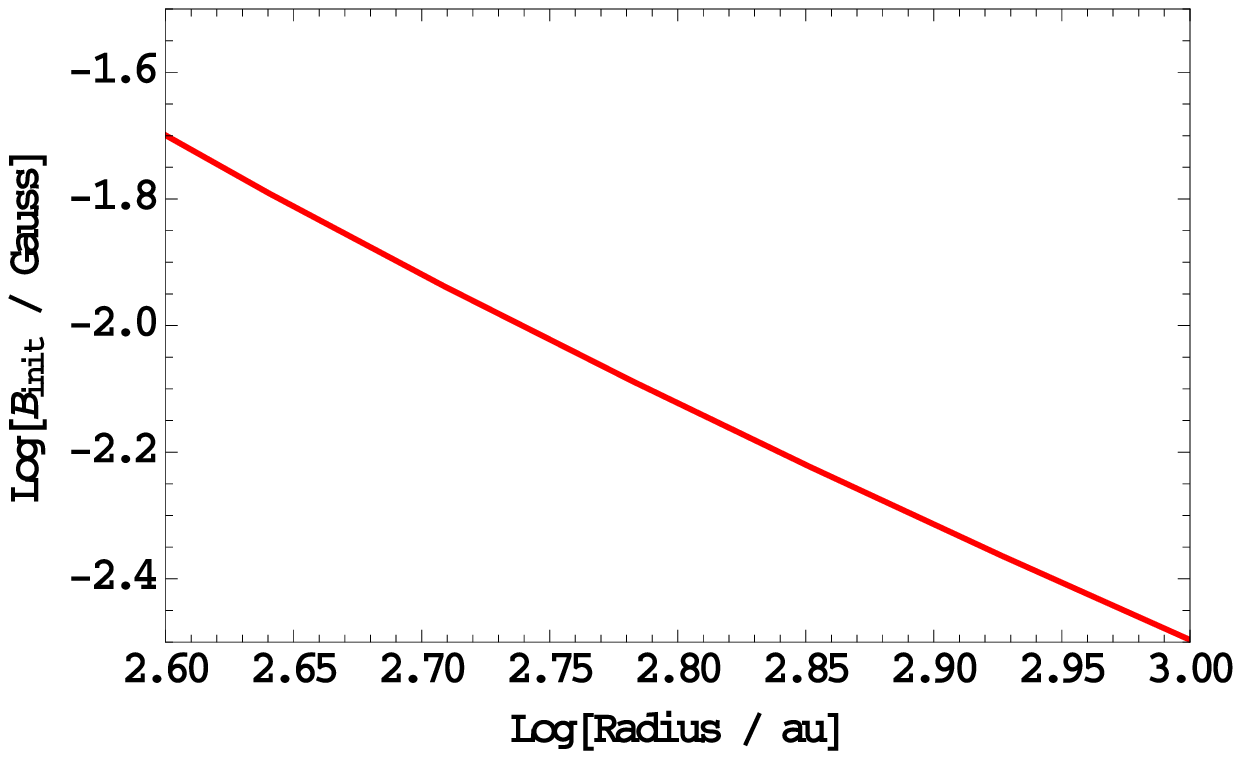}
\end{minipage}  \\
\begin{minipage}{6cm}
\vspace{0.2cm}
\includegraphics[scale=0.7]{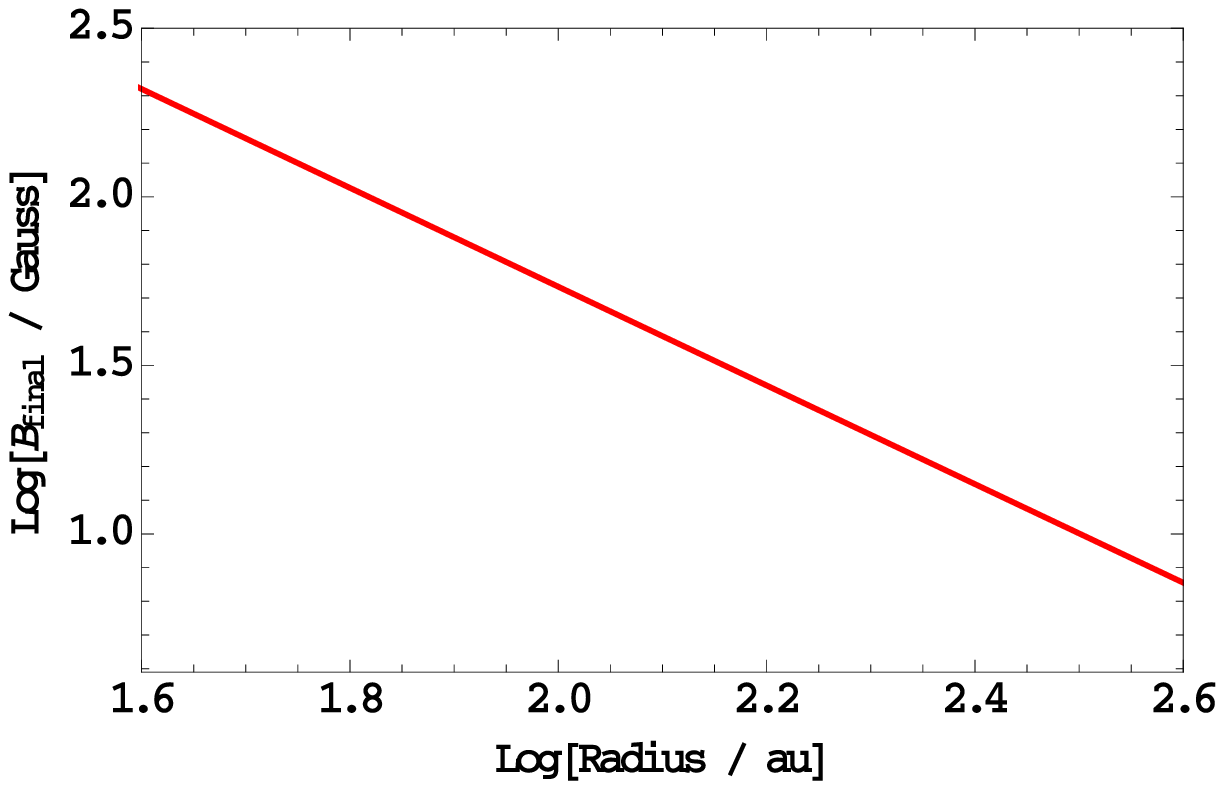}
\end{minipage} &
\begin{minipage}{6cm}
\hspace{2.3cm}
\includegraphics[scale=0.7]{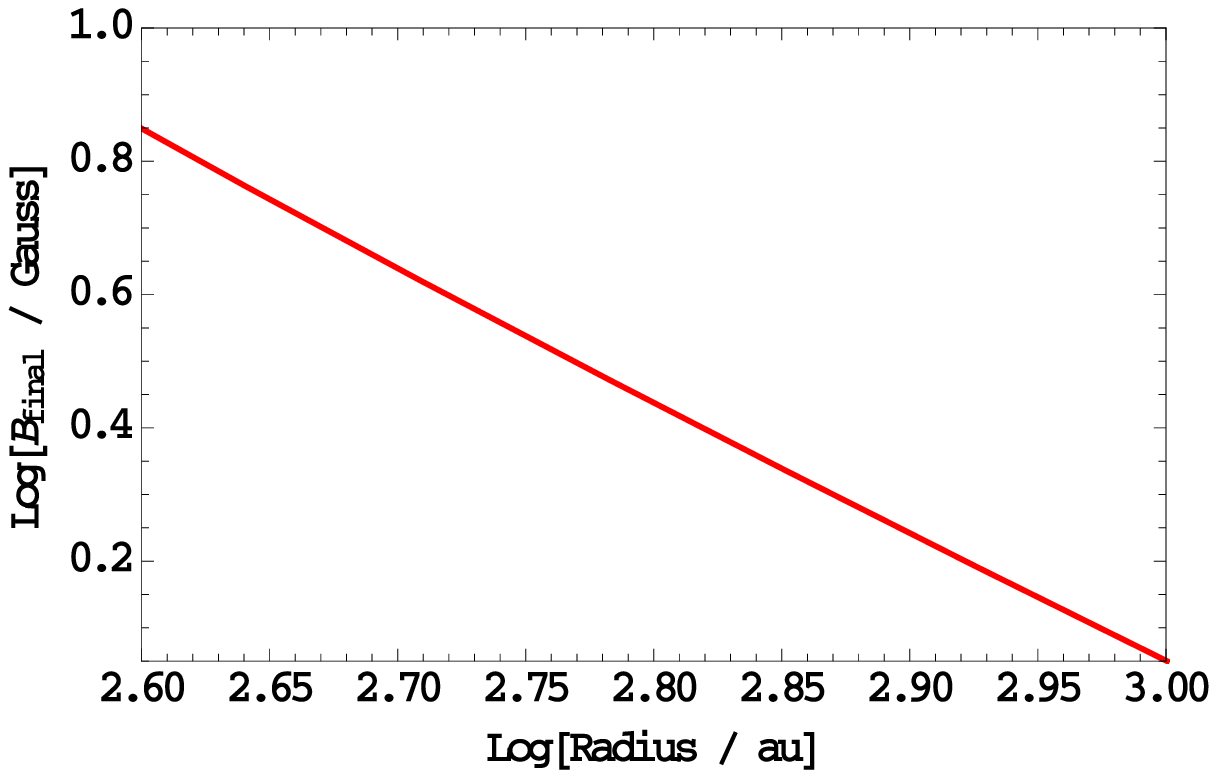}
\end{minipage}\\
\begin{minipage}{6cm}
\vspace{0.2cm}
\includegraphics[scale=0.7]{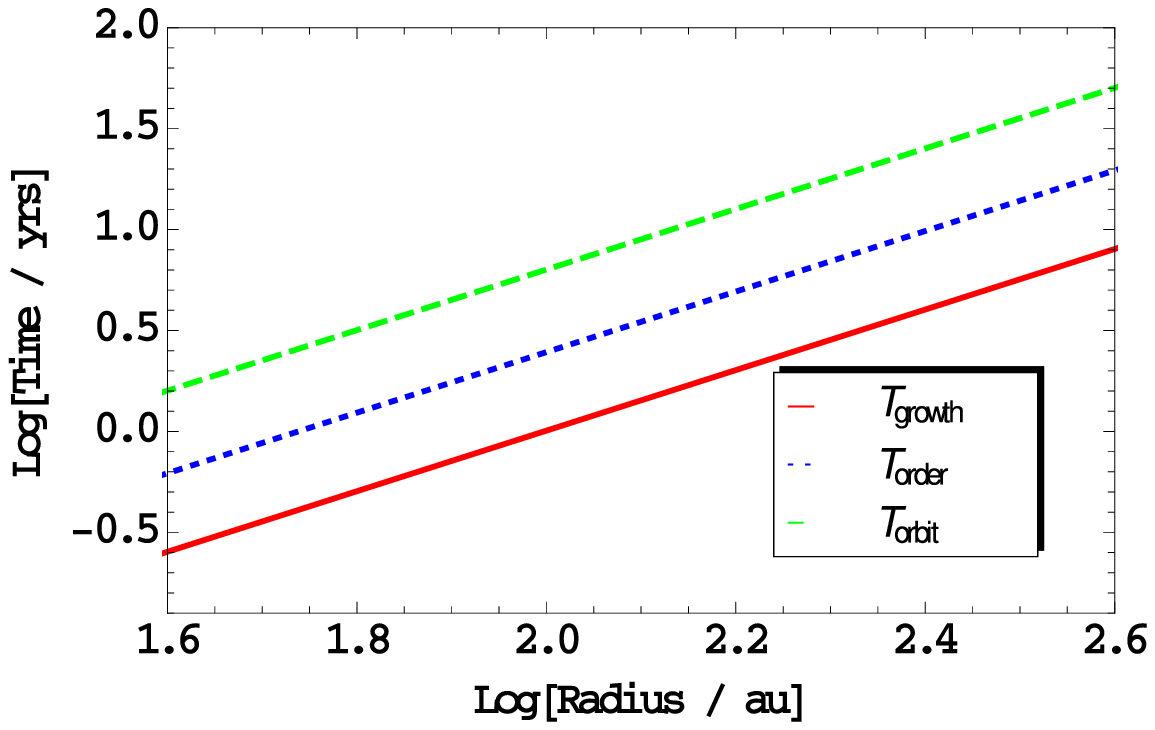}
\end{minipage} &
\begin{minipage}{6cm}
\hspace{2.3cm}
\includegraphics[scale=0.7]{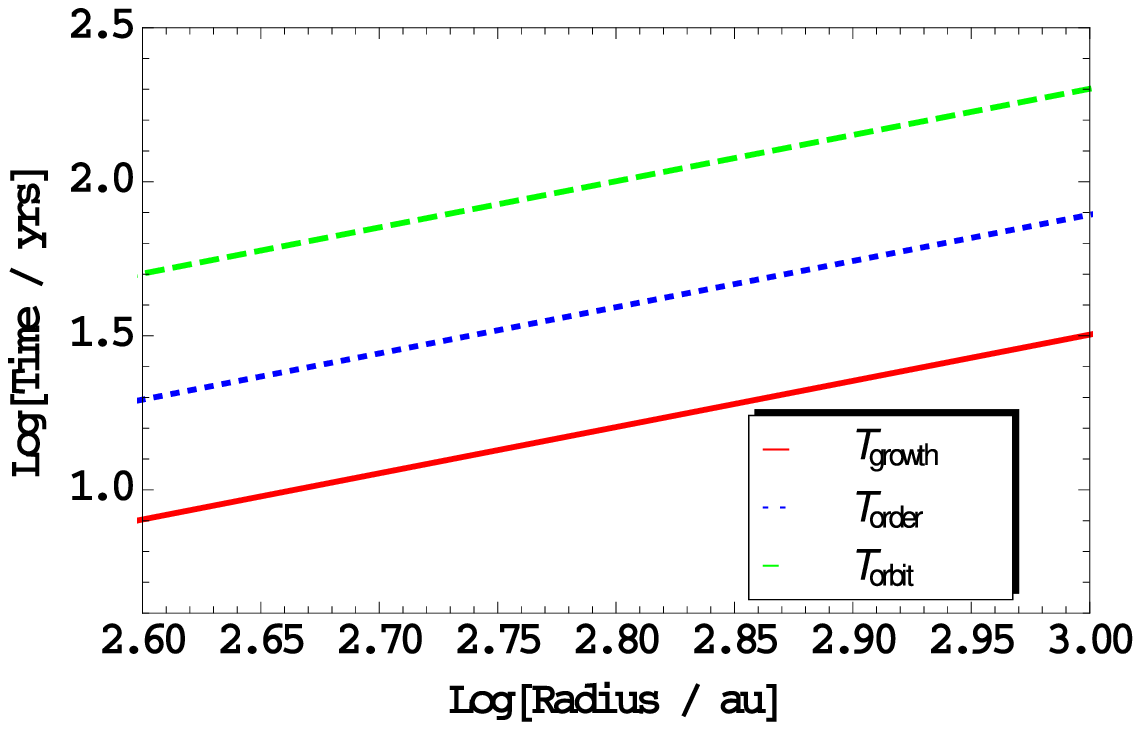}
\end{minipage}
\end{tabular}
\caption{The initial magnetic field strength $B_{init}$ generated by the small-scale dynamo (the top panel) and the final magnetic field strength $B_f$ generated by the $\alpha- \Omega$ dynamo (mid panel) for a $\rm 10^5~M_{\odot}$ star  with an accretion rate of $10^{-1}$~M$_\odot$~yr$^{-1}$. On the bottom panel, we show the orbital timescale, the growth time of the $\alpha-\Omega$ dynamo and the ordering timescale. The left panels show the atomic cooling regime in the inner parts of the disk, while the right panels show the $\rm H_2$ cooling regime on larger scales.}
\label{fig5}
\end{figure*}

The expected magnetic field strengths produced by the dynamo processes in the disk are then given in Fig.~\ref{fig5}. On scales corresponding to the disk radius, the small-scale dynamo contributes an initial field strength $B_{init}$ ranging from $10^{-2.5}$~G at $R=1000$~AU up to $10^{-0.8}$~G at $R \sim39$~AU. The latter is further enhanced via the $\alpha-\Omega$ dynamo, leading to a final field strength $B_f$ of $1-200$~G. The ratio between the initial and final field strength thus varies between $\sim1000$ on large radii and $\sim1260$ on smaller scales, corresponding to $\sim 7$ e-folding times. Due to the considerably larger mass of the central object, the characteristic timescales are significantly reduced compared to the previous scenario, with a growth time of $\sim30$~yrs on scales of $300$~AU. The latter is significantly smaller than the characteristic timescale for accretion, $t_{acc}=M_*/\dot{M}\sim10^6$~yrs, so one can  assume that saturation has occurred on scales of at least $1000$~AU.

\begin{figure*}
\hspace{-6.0cm}
\centering
\begin{tabular}{c c}
\begin{minipage}{6cm}
\vspace{0.2cm}
\includegraphics[scale=0.7]{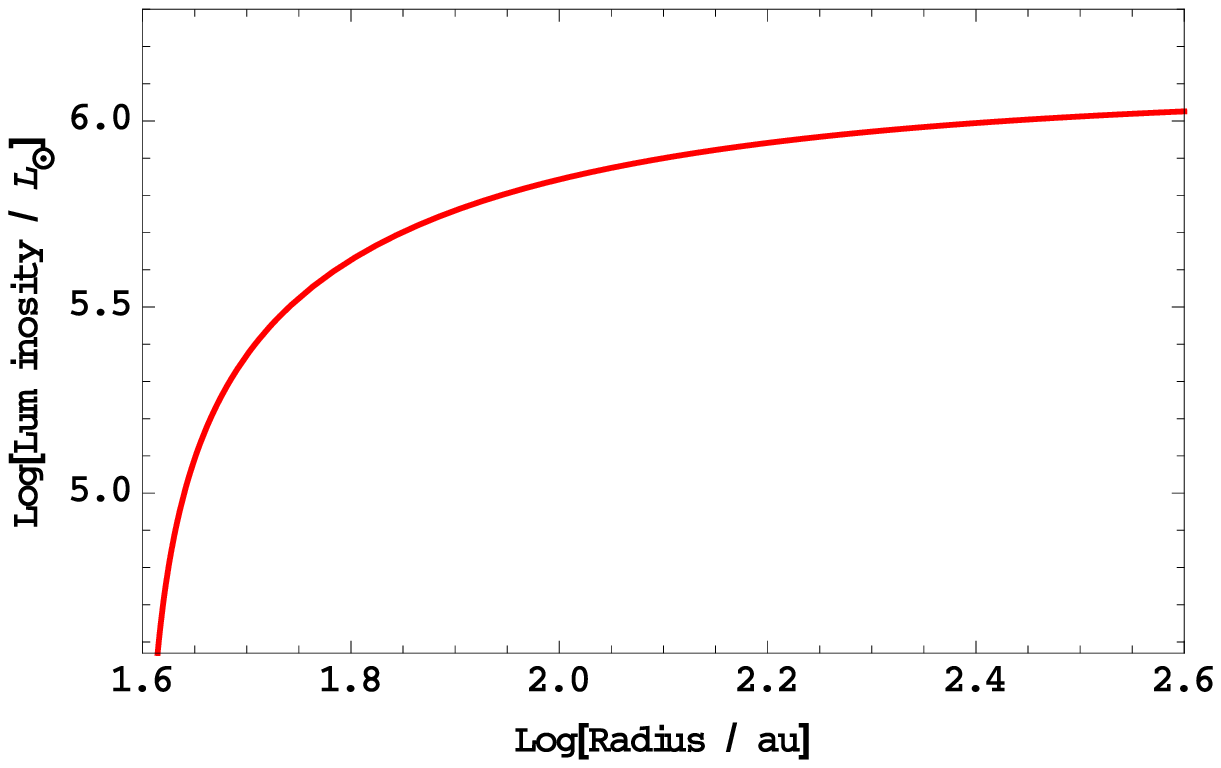}
\end{minipage} &
\begin{minipage}{6cm}
\hspace{2.3cm}
\includegraphics[scale=0.7]{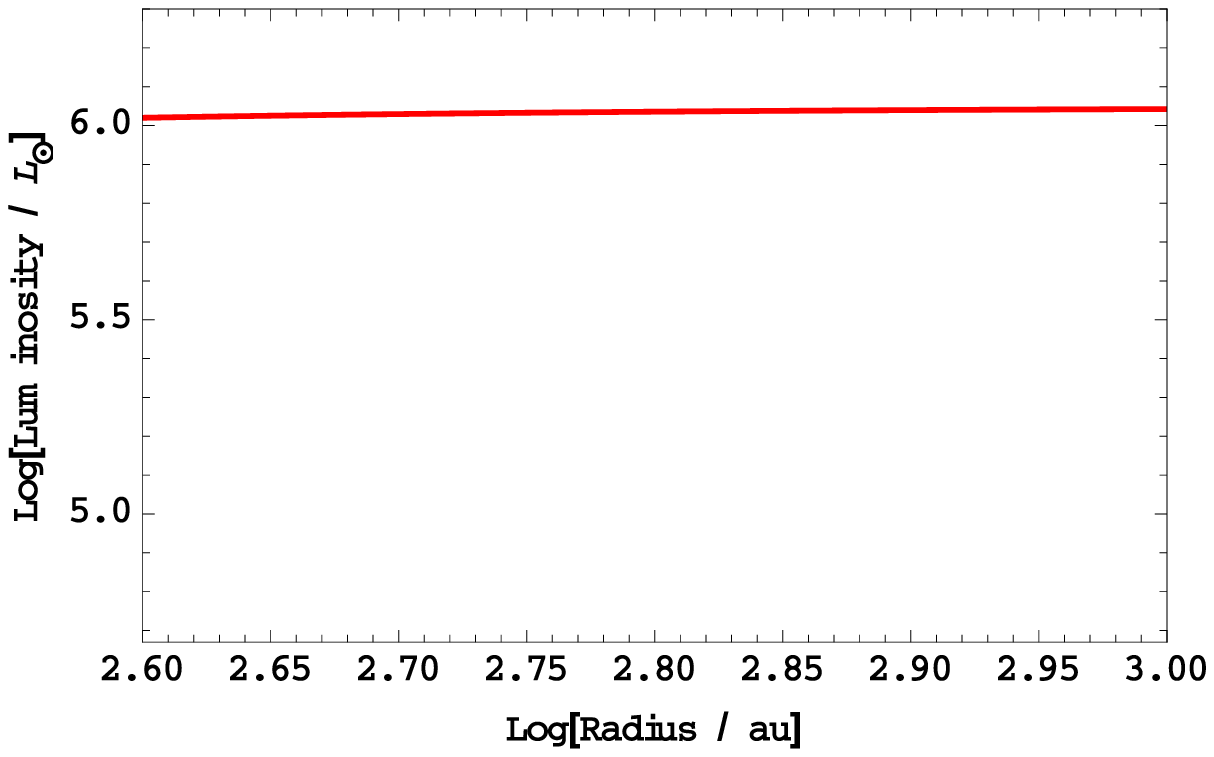}
\end{minipage} \\
\begin{minipage}{6cm}
\vspace{0.2cm}
\includegraphics[scale=0.7]{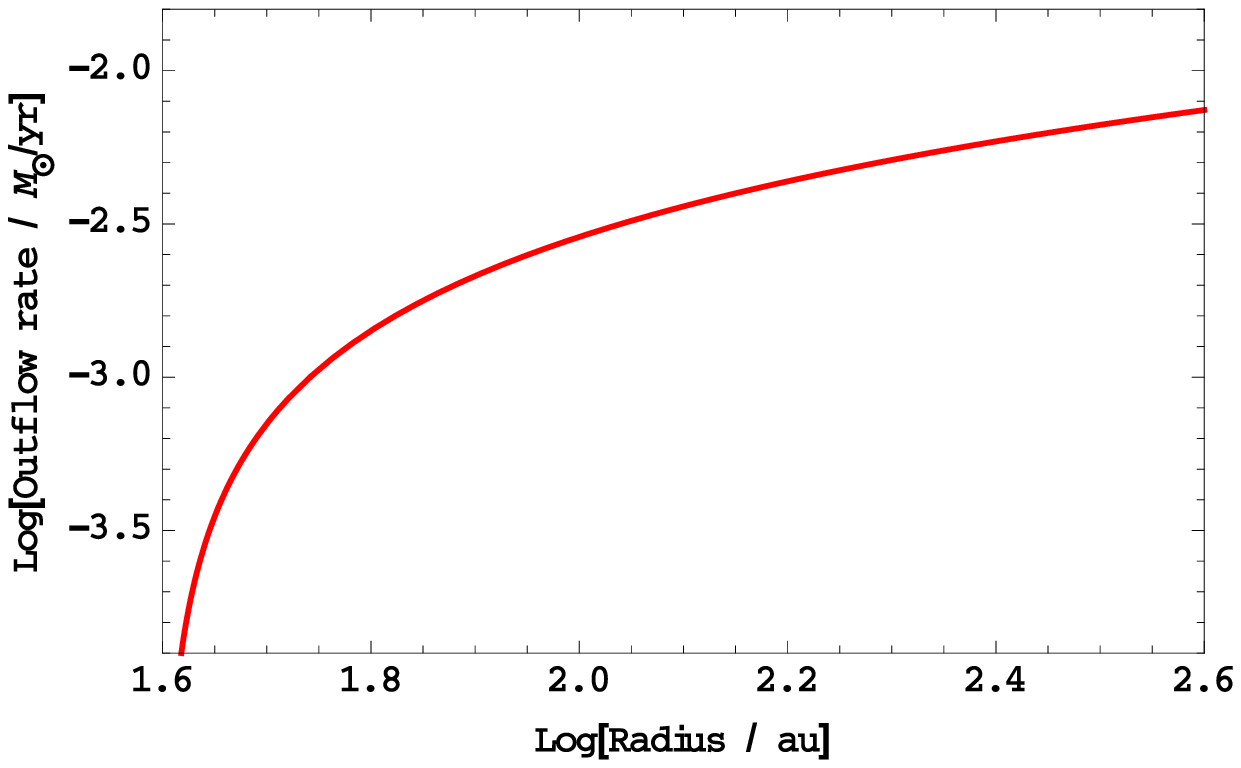}
\end{minipage} &
\begin{minipage}{6cm}
\hspace{2.3cm}
\includegraphics[scale=0.7]{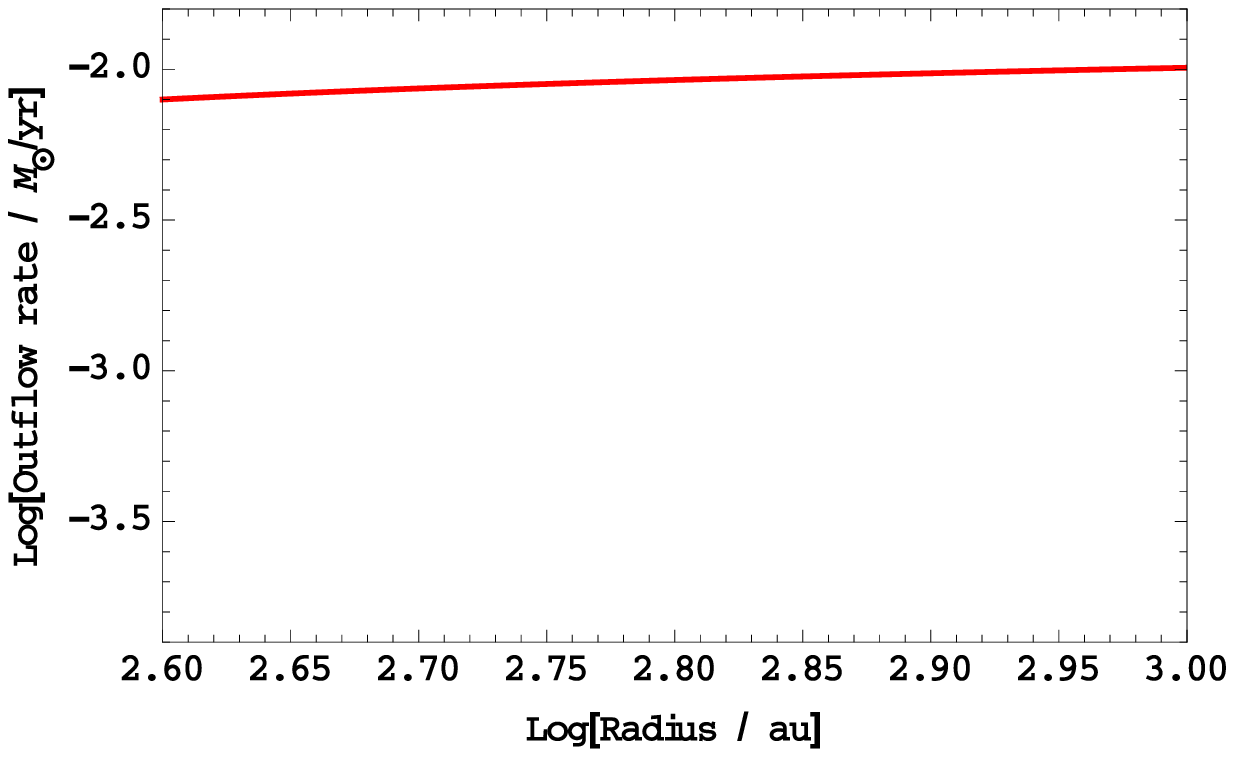}
\end{minipage}
\end{tabular}
\caption{Expected properties of the magnetic outflow for a $10^5$~M$_\odot$ star  with an accretion rate of $10^{-1}$~M$_\odot$~yr$^{-1}$. The magnetic luminosity  is depicted in the top panel while the bottom panel shows the mass outflow rate within a given radius $R$. The left panels represent the atomic cooling regime, while the right panels show the regime of $\rm H_2$ line cooling.}
\label{fig6}
\end{figure*}

The resulting properties of the magnetic outflow are finally shown in Fig.~\ref{fig6}. Both the magnetic luminosity and the mass outflow rate are  dominated by the inner parts of the disk, in this case corresponding to the inner $400$~AU where the gas is atomic. The outflow luminosity at that point corresponds to roughly $10^{6.0}$~L$_\odot$, and the mass outflow rate to $10^{-2.1}$~M$_\odot$~yr$^{-1}$. The outflow therefore does not strongly reduce the accretion onto the central object, but provides a potentially strong and powerful feedback mechanism towards larger scales with a characteristic velocity of $\sim 1200$~km/s.  Especially as the radiative feedback resulting from rapidly accreting supermassive stars is considered to be weak due to their large extended atmospheres \citep[][]{Hosokawa2013, Schleicher13}, the magnetic outflow can provide a relevant mechanism for feedback on larger scales, including the magnetization of the interstellar and intergalactic medium.

The typical lifetime of a supermassive star is about 1 Myrs and is much longer than all other times scales shown in figure \ref{fig5}.  The jet is expected to reach a steady state in a few disk rotation time scales, i.e. $\sim$ 100 yrs \citep{Kudoh2003}. The expected jet velocity is  about 1200 km/s, for which the jet can travel about 1.2 kpc within the star's life. We therefore expect that the jet can have a significant impact on the center of the host galaxy and may affect the subsequent accretion onto the resulting black hole.

\section {Discussion \& conclusions}\label{discussion}

In this paper, we have considered the evolution of magnetic fields after the formation of a self-gravitating primordial disk around the central protostar. While previous studies have shown that a strong tangled magnetic field can be produced via the small-scale dynamo \citep[e.g.][]{Schleicher2010Ma, Sur2010, Schobera, Turk2012, Latif13}, we have explored how the resulting magnetic field can subsequently grow via the $\alpha$-$\Omega$ dynamo in a self-gravitating accretion disk  \citep{Inayoshi14, Latif2014Disk}. Using the resulting magnetic field structure, we have  estimated the expected jet properties and calculated the magnetic luminosity as well as the mass outflow rate. Our model has been applied to two characteristic cases: A conventional $10$~M$_\odot$ Pop.~III star with a typical accretion rate of $10^{-3}$~M$_\odot$~yr$^{-1}$, as well as a $10^5$~M$_\odot$ supermassive star with an accretion rate of $10^{-1}$~M$_\odot$~yr$^{-1}$.

For the conventional Pop.~III star, our results show that the $\alpha$-$\Omega$ dynamo  produces coherent magnetic fields within the protostellar disk at least out to radii of $100$~AU. The resulting strength of the coherent field ranges from $\sim0.1$~G on this particular scale up to $\sim250$~G on a scale of $1$~AU. The predominant contribution to the jet comes from the central $10$~AU, producing a jet luminosity of $\sim 100$~L$_\odot$ and a mass outflow rate of $10^{-3.7}$~M$_\odot$~yr$^{-1}$, implying a negligible mass loss compared to the accretion rate of $10^{-3}$~M$_\odot$~yr$^{-1}$. 

However, while such Pop.~III stars are expected to form in minihalos with characteristic masses of $10^6$~M$_\odot$ and typical escape velocities of $\sim10$~km/s \citep[][]{Abel2002,Bromm2002, Yoshida08}, the characteristic velocity of the jet is about $\sim 60$~km/s, implying that the jet can potentially escape. We estimate that typical timescales for the jet to leave the halo correspond to about $5$~million years. During that time, the jet can provide relevant feedback onto the halo itself, and contribute to the magnetization of the surrounding environment. Particularly,  in dense cosmological environments consisting of several minihalos, the latter may contribute to the magnetization of the intergalactic material.

For the supermassive star, the dynamo amplification is even more efficient due to the shorter characteristic timescales in the stellar environment, implying an efficient magnetization of the disk at least out to radii of $1000$~AU. The resulting magnetic field strengths range from $\sim1$~G on that scale up to $200$~G on scales of $39$~AU. The outflow is predominantly driven within the central $300$~AU, with a jet luminosity of about $10^{6.0}$~L$_\odot$ and a mass outflow rate of $10^{-2.1}$~M$_\odot$~yr$^{-1}$. Also in this case, the outflow is negligible compared to the accretion rate, but can provide a strong mechanical feedback onto the surrounding gas. As previous studies indicated that the radiative feedback of rapidly accreting supermassive stars is rather weak  \citep[][]{Hosokawa2013, Schleicher13}, it is particularly relevant if feedback occurs in the form of jets and outflows. The latter  provides a mechanism to connect the central object with the evolution on larger scales already during the early stages of AGN formation. We  note here that the expected outflow velocity is of the order $1200$~km/s, and therefore considerably larger than the escape velocity of the dark matter halo. Within the lifetime of the star, the jet can therefore propagate over a distance of 1.2 kpc and significantly affect the central region of the host galaxy. The latter is particularly relevant for subsequent accretion onto the central black hole. However, the formation sites of supermassive stars are expected to be quite rare, as they require rather  special conditions to form. The estimates of the number density of supermassive stars suggest that  a few such objects are expected to form  per $\rm Gpc^{-3}$ \citep{Djikstra2014,Habouzit2015}. 

Our results show that magnetic feedback has occured already during the formation of the first generations of stars, with typical mass outflow rates between $1-10\%$ of the mass accretion rate. We expect similar results to hold not only in the primordial case, but even after metal enrichment, for instance in protostellar disk where dust cooling is significant \citep{Tanaka14}. As recently shown by \citet{Susa15}, the coupling between the magnetic field and the gas is rather efficient for a large range of metallicities and densities, and we expect the dynamo mechanism to operate independently of the metallicity of the forming object. We should however note that,  in the case of low-mass objects, the characteristic orbital and dynamo timescales are becoming longer. At the same time, it is also expected that the accretion rate decreases with increasing metallicity, implying that also more time is available for the dynamo to operate. The interplay between such dynamo operation, metal cooling and gas accretion should therefore be studied in greater detail. As a general result, it appears however plausible that magnetized jets and outflows can be expected already from the first generations of protostars, which may act as relevant sources of turbulence at low metallicity in a very similar fashion as in the present-day interstellar medium \citep[e.g.][]{Wang10}.

Beyond the general population of high-redshift stars, it is of particular interest to  explore the consequence of magnetic feedback resulting from supermassive stars, which have been predicted to form in a number of cosmological simulations  \citep{Latif2013d,Regan2014,Latif2014ApJ}. The jets from such supermassive stars provide strong feedback, and can therefore have a substantial influence on the evolution of their host galaxy. Similar to the feedback from supermassive black holes, such jets can potentially suppress star formation in part of the halo \citep{Maiolino12}, or temporarily enhance star formation by compressing the pre-existing clumps \citep{Silk13}. We therefore propose that the resulting feedback mechanism should be included in future simulations studying the formation of the first active galactic nuclei.
%



We have assumed here that turbulence is mainly driven by gravitational instabilities as shown by high resolution cosmological simulations \citep{Turk2012, Latif13}. We note that numerical simulations often underestimate the amount of turbulence during gravitational collapse, as turbulent eddies need to be resolved with at least $32-64$ cells depending on the magneto-hydrodynamical scheme \citep{Sur2010, Federrath11a, Turk2012, Latif13}. Results from such studies therefore need to be regarded as a lower limit. The presence of the MRI may further help in driving the turbulence, and it is even conceivable that both instabilities are operational at the same time \citep{Fromang04}. 

Our model combines various astrophysical processes such as the amplification of magnetic seed fields via the small scale dynamo and the $\alpha - \Omega$ mechanism. We assume that the latter provides a large-scale component of the magnetic field, and that the available Poynting flux will drive a magnetic outflow, as it is well-known in the context of protostars and black holes. A complete modeling of all relevant processes is still difficult under primordial conditions, as the formation of primordial protostellar accretion disks can only be modeled since a few years \citep[see e.g.][]{Clark11, Greif12}. However, the presence of the small-scale dynamo during gravitational collapse has already been confirmed by numerical simulations \citep{Sur2010, Federrath11a, Turk2012, Latif13}, and subsequent studies may now address the further evolution after the formation of a disk in greater detail.

The assumption of an initially turbulent magnetic field has been motivated by the efficient amplification of magnetic fields via the small-scale dynamo, which does not require the existence of a disk. The magneto-rotational instability (MRI) may become operational in the protostellar disks forming around Pop III and supermassive stars, where it can contribute to the generation of turbulence and the amplification of magnetic fields \citep{Silk2006}.

\section*{Acknowledgments}
This project has received funding from the European Union's Horizon 2020 research and innovation programme under the Marie Sklodowska-Curie grant  agreement N$^o$ 656428.  We thank the anonymous referee for his/her feedback which helped us to improve the manuscript.

\bibliography{disk.bib}

\begin{thebibliography}{112}
\expandafter\ifx\csname natexlab\endcsname\relax\def\natexlab#1{#1}\fi

\bibitem[{{Abel} {et~al.}(2002){Abel}, {Bryan}, \& {Norman}}]{Abel2002}
{Abel}, T., {Bryan}, G.~L., \& {Norman}, M.~L. 2002, Science, 295, 93

\bibitem[{{Arlen} {et~al.}(2014){Arlen}, {Vassilev}, {Weisgarber}, {Wakely}, \&
  {Yusef Shafi}}]{ARLEN2014}
{Arlen}, T.~C., {Vassilev}, V.~V., {Weisgarber}, T., {Wakely}, S.~P., \& {Yusef
  Shafi}, S. 2014, \apj, 796, 18

\bibitem[{{Arshakian} {et~al.}(2009){Arshakian}, {Beck}, {Krause}, \&
  {Sokoloff}}]{Arshakian2009}
{Arshakian}, T.~G., {Beck}, R., {Krause}, M., \& {Sokoloff}, D. 2009, \aap,
  494, 21

\bibitem[{{Balbus} \& {Hawley}(1998)}]{Balbus1998}
{Balbus}, S.~A. \& {Hawley}, J.~F. 1998, Reviews of Modern Physics, 70, 1

\bibitem[{{Balsara} \& {Kim}(2005)}]{Balsara05}
{Balsara}, D.~S. \& {Kim}, J. 2005, \apj, 634, 390

\bibitem[{{Balsara} {et~al.}(2004){Balsara}, {Kim}, {Mac Low}, \&
  {Mathews}}]{Balsara2004}
{Balsara}, D.~S., {Kim}, J., {Mac Low}, M.-M., \& {Mathews}, G.~J. 2004, \apj,
  617, 339

\bibitem[{{Banerjee} \& {Pudritz}(2006)}]{Banerjee06}
{Banerjee}, R. \& {Pudritz}, R.~E. 2006, \apj, 641, 949

\bibitem[{{Banerjee} {et~al.}(2009){Banerjee}, {V{\'a}zquez-Semadeni},
  {Hennebelle}, \& {Klessen}}]{Banerjee09}
{Banerjee}, R., {V{\'a}zquez-Semadeni}, E., {Hennebelle}, P., \& {Klessen},
  R.~S. 2009, \mnras, 398, 1082

\bibitem[{{Beck} {et~al.}(1996){Beck}, {Brandenburg}, {Moss}, {Shukurov}, \&
  {Sokoloff}}]{Beck96}
{Beck}, R., {Brandenburg}, A., {Moss}, D., {Shukurov}, A., \& {Sokoloff}, D.
  1996, \araa, 34, 155

\bibitem[{{Beresnyak}(2012)}]{Beresnyak12}
{Beresnyak}, A. 2012, Physical Review Letters, 108, 035002

\bibitem[{{Bernet} {et~al.}(2008){Bernet}, {Miniati}, {Lilly}, {Kronberg}, \&
  {Dessauges-Zavadsky}}]{Bernet08}
{Bernet}, M.~L., {Miniati}, F., {Lilly}, S.~J., {Kronberg}, P.~P., \&
  {Dessauges-Zavadsky}, M. 2008, \nat, 454, 302

\bibitem[{{Biermann}(1950)}]{Biermann1950}
{Biermann}, L. 1950, Zeitschrift Naturforschung Teil A, 5, 65

\bibitem[{{Bovino} {et~al.}(2014){Bovino}, {Latif}, {Grassi}, \&
  {Schleicher}}]{BovinoHD2014}
{Bovino}, S., {Latif}, M.~A., {Grassi}, T., \& {Schleicher}, D.~R.~G. 2014,
  \mnras, 441, 2181

\bibitem[{{Bovino} {et~al.}(2013){Bovino}, {Schleicher}, \&
  {Schober}}]{Bovino13}
{Bovino}, S., {Schleicher}, D.~R.~G., \& {Schober}, J. 2013, New Journal of
  Physics, 15, 013055

\bibitem[{{Brandenburg} \& {Subramanian}(2005)}]{Brandenburg05}
{Brandenburg}, A. \& {Subramanian}, K. 2005, \physrep, 417, 1

\bibitem[{{Bromm} {et~al.}(2002){Bromm}, {Coppi}, \& {Larson}}]{Bromm2002}
{Bromm}, V., {Coppi}, P.~S., \& {Larson}, R.~B. 2002, \apj, 564, 23

\bibitem[{{Clark} {et~al.}(2011){Clark}, {Glover}, {Smith}, {Greif}, {Klessen},
  \& {Bromm}}]{Clark11}
{Clark}, P.~C., {Glover}, S.~C.~O., {Smith}, R.~J., {et~al.} 2011, Science,
  331, 1040

\bibitem[{{de Souza} \& {Opher}(2010)}]{Souza10}
{de Souza}, R.~S. \& {Opher}, R. 2010, \prd, 81, 067301

\bibitem[{{Dermer} {et~al.}(2011){Dermer}, {Cavadini}, {Razzaque}, {Finke},
  {Chiang}, \& {Lott}}]{Dermer2011}
{Dermer}, C.~D., {Cavadini}, M., {Razzaque}, S., {et~al.} 2011, \apjl, 733, L21

\bibitem[{{Dijkstra} {et~al.}(2014){Dijkstra}, {Ferrara}, \&
  {Mesinger}}]{Djikstra2014}
{Dijkstra}, M., {Ferrara}, A., \& {Mesinger}, A. 2014, \mnras, 442, 2036

\bibitem[{{Dolag} {et~al.}(2011){Dolag}, {Kachelriess}, {Ostapchenko}, \&
  {Tom{\`a}s}}]{Dolag11}
{Dolag}, K., {Kachelriess}, M., {Ostapchenko}, S., \& {Tom{\`a}s}, R. 2011,
  \apjl, 727, L4

\bibitem[{{Elmegreen} \& {Burkert}(2010)}]{Elmegreen10}
{Elmegreen}, B.~G. \& {Burkert}, A. 2010, ApJ, 712, 294

\bibitem[{{Federrath} {et~al.}(2011{\natexlab{a}}){Federrath}, {Chabrier},
  {Schober}, {Banerjee}, {Klessen}, \& {Schleicher}}]{Federrath11b}
{Federrath}, C., {Chabrier}, G., {Schober}, J., {et~al.} 2011{\natexlab{a}},
  Physical Review Letters, 107, 114504

\bibitem[{{Federrath} {et~al.}(2014){Federrath}, {Schober}, {Bovino}, \&
  {Schleicher}}]{Federrath14}
{Federrath}, C., {Schober}, J., {Bovino}, S., \& {Schleicher}, D.~R.~G. 2014,
  \apjl, 797, L19

\bibitem[{{Federrath} {et~al.}(2011{\natexlab{b}}){Federrath}, {Sur},
  {Schleicher}, {Banerjee}, \& {Klessen}}]{Federrath11a}
{Federrath}, C., {Sur}, S., {Schleicher}, D.~R.~G., {Banerjee}, R., \&
  {Klessen}, R.~S. 2011{\natexlab{b}}, \apj, 731, 62

\bibitem[{{Fendt}(2009)}]{Fendt09}
{Fendt}, C. 2009, \apj, 692, 346

\bibitem[{{Ferrara} {et~al.}(2014){Ferrara}, {Salvadori}, {Yue}, \&
  {Schleicher}}]{Ferrara14}
{Ferrara}, A., {Salvadori}, S., {Yue}, B., \& {Schleicher}, D. 2014, \mnras,
  443, 2410

\bibitem[{{Fromang} {et~al.}(2004){Fromang}, {Balbus}, {Terquem}, \& {De
  Villiers}}]{Fromang04}
{Fromang}, S., {Balbus}, S.~A., {Terquem}, C., \& {De Villiers}, J.-P. 2004,
  \apj, 616, 364

\bibitem[{{Gao} {et~al.}(2010){Gao}, {Theuns}, {Frenk}, {Jenkins}, {Helly},
  {Navarro}, {Springel}, \& {White}}]{Gao2010}
{Gao}, L., {Theuns}, T., {Frenk}, C.~S., {et~al.} 2010, \mnras, 403, 1283

\bibitem[{{Glover} \& {Savin}(2009)}]{Glover09}
{Glover}, S.~C.~O. \& {Savin}, D.~W. 2009, \mnras, 393, 911

\bibitem[{{Grasso} \& {Rubinstein}(2001)}]{Grasso01}
{Grasso}, D. \& {Rubinstein}, H.~R. 2001, \physrep, 348, 163

\bibitem[{{Greif} {et~al.}(2012){Greif}, {Bromm}, {Clark}, {Glover}, {Smith},
  {Klessen}, {Yoshida}, \& {Springel}}]{Greif12}
{Greif}, T.~H., {Bromm}, V., {Clark}, P.~C., {et~al.} 2012, \mnras, 424, 399

\bibitem[{{Gressel} {et~al.}(2008){Gressel}, {Elstner}, {Ziegler}, \&
  {R{\"u}diger}}]{Gressel2008}
{Gressel}, O., {Elstner}, D., {Ziegler}, U., \& {R{\"u}diger}, G. 2008, \aap,
  486, L35

\bibitem[{{Habouzit} {et~al.}(2015){Habouzit}, {Volonteri}, {Latif},
  {Nishimichi}, {Peirani}, {Dubois}, {Mamon}, {Silk}, \&
  {Chevallard}}]{Habouzit2015}
{Habouzit}, M., {Volonteri}, M., {Latif}, M., {et~al.} 2015, ArXiv
  e-prints:1507.05971

\bibitem[{{Hammond} {et~al.}(2012){Hammond}, {Robishaw}, \&
  {Gaensler}}]{2012arXiv1209.1438H}
{Hammond}, A.~M., {Robishaw}, T., \& {Gaensler}, B.~M. 2012, ArXiv e-prints
  1209.1438

\bibitem[{{Hawley} \& {Balbus}(1991)}]{Hawley91}
{Hawley}, J.~F. \& {Balbus}, S.~A. 1991, \apj, 376, 223

\bibitem[{{Hennebelle} \& {Fromang}(2008)}]{Hennebelle08}
{Hennebelle}, P. \& {Fromang}, S. 2008, \aap, 477, 9

\bibitem[{{Hennebelle} \& {Teyssier}(2008)}]{Hennebelle08b}
{Hennebelle}, P. \& {Teyssier}, R. 2008, \aap, 477, 25

\bibitem[{{Hogan}(1983)}]{Hogan1983}
{Hogan}, C.~J. 1983, Physical Review Letters, 51, 1488

\bibitem[{{Hosokawa} {et~al.}(2013){Hosokawa}, {Yorke}, {Inayoshi}, {Omukai},
  \& {Yoshida}}]{Hosokawa2013}
{Hosokawa}, T., {Yorke}, H.~W., {Inayoshi}, K., {Omukai}, K., \& {Yoshida}, N.
  2013, \apj, 778, 178

\bibitem[{{Inayoshi} \& {Haiman}(2014)}]{Inayoshi14}
{Inayoshi}, K. \& {Haiman}, Z. 2014, \mnras, 445, 1549

\bibitem[{{Ivison} \& {et al.}(2010)}]{Ivison10}
{Ivison}, R.~J. \& {et al.} 2010, \aap, 518, L31

\bibitem[{{Kazantsev}(1968)}]{Kazantsev68}
{Kazantsev}, A.~P. 1968, Soviet Journal of Experimental and Theoretical
  Physics, 26, 1031

\bibitem[{{Klessen} \& {Hennebelle}(2010)}]{Klessen10}
{Klessen}, R.~S. \& {Hennebelle}, P. 2010, A\&A, 520, A17

\bibitem[{{Konigl} \& {Pudritz}(2000)}]{Konigl2000}
{Konigl}, A. \& {Pudritz}, R.~E. 2000, Protostars and Planets IV, 759

\bibitem[{{Krause} \& {Raedler}(1980)}]{KrauseBook1980}
{Krause}, F. \& {Raedler}, K.~H. 1980, {Mean-field magnetohydrodynamics and
  dynamo theory} (Oxford: Pergamon Press, 1980)

\bibitem[{{Kronberg} {et~al.}(2008){Kronberg}, {Bernet}, {Miniati}, {Lilly},
  {Short}, \& {Higdon}}]{Kronberg08}
{Kronberg}, P.~P., {Bernet}, M.~L., {Miniati}, F., {et~al.} 2008, \apj, 676, 70

\bibitem[{{Kudoh} {et~al.}(2003){Kudoh}, {Matsumoto}, \& {Shibata}}]{Kudoh2003}
{Kudoh}, T., {Matsumoto}, R., \& {Shibata}, K. 2003, \apss, 287, 99

\bibitem[{{Kulsrud}(1999)}]{Kulsrud1999}
{Kulsrud}, R.~M. 1999, \araa, 37, 37

\bibitem[{{Kulsrud} \& {Zweibel}(2008)}]{Kulsrud2008}
{Kulsrud}, R.~M. \& {Zweibel}, E.~G. 2008, Reports on Progress in Physics, 71,
  046901

\bibitem[{{Latif} \& {Schlei\-cher}(2015{\natexlab{a}})}]{Latif2014Disk}
{Latif}, M.~A. \& {Schlei\-cher}, D.~R.~G. 2015{\natexlab{a}}, \mnras, 449, 77

\bibitem[{{Latif} \& {Schlei\-cher}(2015{\natexlab{b}})}]{Latif2015c}
{Latif}, M.~A. \& {Schlei\-cher}, D.~R.~G. 2015{\natexlab{b}}, ArXiv
  e-prints:1411.5902

\bibitem[{{Latif} {et~al.}(2014{\natexlab{a}}){Latif}, {Schleicher}, {Bovino},
  {Grassi}, \& {Spaans}}]{Latif2014ApJ}
{Latif}, M.~A., {Schleicher}, D.~R.~G., {Bovino}, S., {Grassi}, T., \&
  {Spaans}, M. 2014{\natexlab{a}}, \apj, 792, 78

\bibitem[{{Latif} {et~al.}(2014{\natexlab{b}}){Latif}, {Schleicher}, \&
  {Schmidt}}]{Latif2014Magnetic}
{Latif}, M.~A., {Schleicher}, D.~R.~G., \& {Schmidt}, W. 2014{\natexlab{b}},
  \mnras, 440, 1551

\bibitem[{{Latif} {et~al.}(2013{\natexlab{a}}){Latif}, {Schleicher}, {Schmidt},
  \& {Niemeyer}}]{Latif2013c}
{Latif}, M.~A., {Schleicher}, D.~R.~G., {Schmidt}, W., \& {Niemeyer}, J.
  2013{\natexlab{a}}, \mnras, 433, 1607

\bibitem[{{Latif} {et~al.}(2013{\natexlab{b}}){Latif}, {Schleicher}, {Schmidt},
  \& {Niemeyer}}]{Latif2013a}
{Latif}, M.~A., {Schleicher}, D.~R.~G., {Schmidt}, W., \& {Niemeyer}, J.
  2013{\natexlab{b}}, \mnras, 430, 588

\bibitem[{{Latif} {et~al.}(2013{\natexlab{c}}){Latif}, {Schleicher}, {Schmidt},
  \& {Niemeyer}}]{LatifPopIII13}
{Latif}, M.~A., {Schleicher}, D.~R.~G., {Schmidt}, W., \& {Niemeyer}, J.
  2013{\natexlab{c}}, \apjl, 772, L3

\bibitem[{{Latif} {et~al.}(2013{\natexlab{d}}){Latif}, {Schleicher}, {Schmidt},
  \& {Niemeyer}}]{Latif13}
{Latif}, M.~A., {Schleicher}, D.~R.~G., {Schmidt}, W., \& {Niemeyer}, J.
  2013{\natexlab{d}}, \mnras, 432, 668

\bibitem[{{Latif} {et~al.}(2013{\natexlab{e}}){Latif}, {Schleicher}, {Schmidt},
  \& {Niemeyer}}]{Latif2013d}
{Latif}, M.~A., {Schleicher}, D.~R.~G., {Schmidt}, W., \& {Niemeyer}, J.~C.
  2013{\natexlab{e}}, \mnras, 436, 2989

\bibitem[{{Latif} \& {Volonteri}(2015)}]{LatifVolonteri2015}
{Latif}, M.~A. \& {Volonteri}, M. 2015, \mnras, 452, 1026

\bibitem[{{Lodato}(2007)}]{Lodato2007}
{Lodato}, G. 2007, Nuovo Cimento Rivista Serie, 30, 293

\bibitem[{{Lovelace} {et~al.}(2002){Lovelace}, {Li}, {Koldoba}, {Ustyugova}, \&
  {Romanova}}]{Lovelace2002}
{Lovelace}, R.~V.~E., {Li}, H., {Koldoba}, A.~V., {Ustyugova}, G.~V., \&
  {Romanova}, M.~M. 2002, \apj, 572, 445

\bibitem[{{Machida} \& {Doi}(2013)}]{Machida2013}
{Machida}, M.~N. \& {Doi}, K. 2013, \mnras, 435, 3283

\bibitem[{{Machida} {et~al.}(2008){Machida}, {Matsumoto}, \&
  {Inutsuka}}]{Machida08}
{Machida}, M.~N., {Matsumoto}, T., \& {Inutsuka}, S.-i. 2008, \apj, 685, 690

\bibitem[{{Machida} {et~al.}(2006){Machida}, {Omukai}, {Matsumoto}, \&
  {Inutsuka}}]{Machida2006}
{Machida}, M.~N., {Omukai}, K., {Matsumoto}, T., \& {Inutsuka}, S.-i. 2006,
  \apjl, 647, L1

\bibitem[{{Magnelli} \& {et al.}(2015)}]{Magnelli15}
{Magnelli}, B. \& {et al.} 2015, \aap, 573, A45

\bibitem[{{Maiolino} {et~al.}(2012){Maiolino}, {Gallerani}, {Neri}, {Cicone},
  {Ferrara}, {Genzel}, {Lutz}, {Sturm}, {Tacconi}, {Walter}, {Feruglio},
  {Fiore}, \& {Piconcelli}}]{Maiolino12}
{Maiolino}, R., {Gallerani}, S., {Neri}, R., {et~al.} 2012, \mnras, 425, L66

\bibitem[{{Maki} \& {Susa}(2004)}]{Maki04}
{Maki}, H. \& {Susa}, H. 2004, \apj, 609, 467

\bibitem[{{Maki} \& {Susa}(2007)}]{Maki07}
{Maki}, H. \& {Susa}, H. 2007, \pasj, 59, 787

\bibitem[{{Moss} {et~al.}(2013){Moss}, {Beck}, {Sokoloff}, {Stepanov},
  {Krause}, \& {Arshakian}}]{Moss2013}
{Moss}, D., {Beck}, R., {Sokoloff}, D., {et~al.} 2013, \aap, 556, A147

\bibitem[{{Moss} {et~al.}(1998){Moss}, {Shukurov}, \& {Sokoloff}}]{Moss1998}
{Moss}, D., {Shukurov}, A., \& {Sokoloff}, D. 1998, Geophysical and
  Astrophysical Fluid Dynamics, 89, 285

\bibitem[{{Moss} {et~al.}(2012){Moss}, {Stepanov}, {Arshakian}, {Beck},
  {Krause}, \& {Sokoloff}}]{Moss2012}
{Moss}, D., {Stepanov}, R., {Arshakian}, T.~G., {et~al.} 2012, \aap, 537, A68

\bibitem[{{Pakmor} \& {Springel}(2013)}]{Pakmor2013}
{Pakmor}, R. \& {Springel}, V. 2013, \mnras, 432, 176

\bibitem[{{Parker}(1971)}]{Parker1971}
{Parker}, E.~N. 1971, \apj, 163, 255

\bibitem[{{Peters} {et~al.}(2012){Peters}, {Schleicher}, {Klessen}, {Banerjee},
  {Federrath}, {Smith}, \& {Sur}}]{Peters12}
{Peters}, T., {Schleicher}, D.~R.~G., {Klessen}, R.~S., {et~al.} 2012, \apjl,
  760, L28

\bibitem[{{Price} \& {Bate}(2008)}]{Price08}
{Price}, D.~J. \& {Bate}, M.~R. 2008, \mnras, 385, 1820

\bibitem[{{Pudritz} \& {Silk}(1989)}]{Pudritz89}
{Pudritz}, R.~E. \& {Silk}, J. 1989, \apj, 342, 650

\bibitem[{{Reed} {et~al.}(2005){Reed}, {Bower}, {Frenk}, {Gao}, {Jenkins},
  {Theuns}, \& {White}}]{Reed2005}
{Reed}, D.~S., {Bower}, R., {Frenk}, C.~S., {et~al.} 2005, \mnras, 363, 393

\bibitem[{{Regan} {et~al.}(2014){Regan}, {Johansson}, \&
  {Haehnelt}}]{Regan2014}
{Regan}, J.~A., {Johansson}, P.~H., \& {Haehnelt}, M.~G. 2014, \mnras, 439,
  1160

\bibitem[{{Ruzmaikin} {et~al.}(1988){Ruzmaikin}, {Sokolov}, \&
  {Shukurov}}]{Ruzmaikin1988}
{Ruzmaikin}, A.~A., {Sokolov}, D.~D., \& {Shukurov}, A.~M., eds. 1988,
  Astrophysics and Space Science Library, Vol. 133, {Magnetic fields of
  galaxies}

\bibitem[{{Ryu} {et~al.}(2012){Ryu}, {Schleicher}, {Treumann}, {Tsagas}, \&
  {Widrow}}]{Ryu2012}
{Ryu}, D., {Schleicher}, D.~R.~G., {Treumann}, R.~A., {Tsagas}, C.~G., \&
  {Widrow}, L.~M. 2012, \ssr, 166, 1

\bibitem[{{Schekochihin} {et~al.}(2002){Schekochihin}, {Cowley}, {Hammett},
  {Maron}, \& {McWilliams}}]{Scheko02}
{Schekochihin}, A.~A., {Cowley}, S.~C., {Hammett}, G.~W., {Maron}, J.~L., \&
  {McWilliams}, J.~C. 2002, New Journal of Physics, 4, 84

\bibitem[{{Schlei\-cher} {et~al.}(2010){Schlei\-cher}, {Banerjee}, {Sur},
  {Arshakian}, {Klessen}, {Beck}, \& {Spaans}}]{Schleicher2010Ma}
{Schlei\-cher}, D.~R.~G., {Banerjee}, R., {Sur}, S., {et~al.} 2010, \aap, 522,
  A115

\bibitem[{{Schleicher} {et~al.}(2013{\natexlab{a}}){Schleicher}, {Palla},
  {Ferrara}, {Galli}, \& {Latif}}]{Schleicher13}
{Schleicher}, D.~R.~G., {Palla}, F., {Ferrara}, A., {Galli}, D., \& {Latif}, M.
  2013{\natexlab{a}}, \aap, 558, A59

\bibitem[{{Schleicher} {et~al.}(2013{\natexlab{b}}){Schleicher}, {Schober},
  {Federrath}, {Bovino}, \& {Schmidt}}]{Schleicher13a}
{Schleicher}, D.~R.~G., {Schober}, J., {Federrath}, C., {Bovino}, S., \&
  {Schmidt}, W. 2013{\natexlab{b}}, New Journal of Physics, 15, 023017

\bibitem[{{Schlickeiser}(2012)}]{Schlickeiser12}
{Schlickeiser}, R. 2012, Physical Review Letters, 109, 261101

\bibitem[{{Schmitt}(1990)}]{Schmitt1990}
{Schmitt}, D. 1990, in Reviews in Modern Astronomy, Vol.~3, Reviews in Modern
  Astronomy, ed. G.~{Klare}, 86--97

\bibitem[{{Schober} {et~al.}(2012{\natexlab{a}}){Schober}, {Schleicher},
  {Federrath}, {Glover}, {Klessen}, \& {Banerjee}}]{Schobera}
{Schober}, J., {Schleicher}, D., {Federrath}, C., {et~al.} 2012{\natexlab{a}},
  \apj, 754, 99

\bibitem[{{Schober} {et~al.}(2012{\natexlab{b}}){Schober}, {Schleicher},
  {Federrath}, {Klessen}, \& {Banerjee}}]{Schoberb}
{Schober}, J., {Schleicher}, D., {Federrath}, C., {Klessen}, R., \& {Banerjee},
  R. 2012{\natexlab{b}}, \pre, 85, 026303

\bibitem[{{Shakura} \& {Sunyaev}(1973)}]{Shakura73}
{Shakura}, N.~I. \& {Sunyaev}, R.~A. 1973, \aap, 24, 337

\bibitem[{{Shu} {et~al.}(2007){Shu}, {Galli}, {Lizano}, {Glassgold}, \&
  {Diamond}}]{Shu07}
{Shu}, F.~H., {Galli}, D., {Lizano}, S., {Glassgold}, A.~E., \& {Diamond},
  P.~H. 2007, \apj, 665, 535

\bibitem[{{Shu} {et~al.}(1995){Shu}, {Najita}, {Ostriker}, \& {Shang}}]{Shu95}
{Shu}, F.~H., {Najita}, J., {Ostriker}, E.~C., \& {Shang}, H. 1995, \apjl, 455,
  L155

\bibitem[{{Shu} {et~al.}(2000){Shu}, {Najita}, {Shang}, \& {Li}}]{Shu2000}
{Shu}, F.~H., {Najita}, J.~R., {Shang}, H., \& {Li}, Z.-Y. 2000, Protostars and
  Planets IV, 789

\bibitem[{{Silk}(2013)}]{Silk13}
{Silk}, J. 2013, \apj, 772, 112

\bibitem[{{Silk} \& {Langer}(2006)}]{Silk2006}
{Silk}, J. \& {Langer}, M. 2006, \mnras, 371, 444

\bibitem[{{Steenbeck} {et~al.}(1966){Steenbeck}, {Krause}, \&
  {R{\"a}dler}}]{Steeback1966}
{Steenbeck}, M., {Krause}, F., \& {R{\"a}dler}, K.-H. 1966, Zeitschrift
  Naturforschung Teil A, 21, 369

\bibitem[{{Sur} {et~al.}(2012){Sur}, {Federrath}, {Schleicher}, {Banerjee}, \&
  {Klessen}}]{Sur12}
{Sur}, S., {Federrath}, C., {Schleicher}, D.~R.~G., {Banerjee}, R., \&
  {Klessen}, R.~S. 2012, \mnras, 423, 3148

\bibitem[{{Sur} {et~al.}(2010){Sur}, {Schleicher}, {Banerjee}, {Federrath}, \&
  {Klessen}}]{Sur2010}
{Sur}, S., {Schleicher}, D.~R.~G., {Banerjee}, R., {Federrath}, C., \&
  {Klessen}, R.~S. 2010, \apjl, 721, L134

\bibitem[{{Susa} {et~al.}(2015){Susa}, {Doi}, \& {Omukai}}]{Susa15}
{Susa}, H., {Doi}, K., \& {Omukai}, K. 2015, \apj, 801, 13

\bibitem[{{Tan} \& {Blackman}(2004)}]{Tan2004}
{Tan}, J.~C. \& {Blackman}, E.~G. 2004, \apj, 603, 401

\bibitem[{{Tanaka} \& {Omukai}(2014)}]{Tanaka14}
{Tanaka}, K.~E.~I. \& {Omukai}, K. 2014, \mnras, 439, 1884

\bibitem[{{Tavecchio} {et~al.}(2011){Tavecchio}, {Ghisellini}, {Bonnoli}, \&
  {Foschini}}]{Tavecchio11}
{Tavecchio}, F., {Ghisellini}, G., {Bonnoli}, G., \& {Foschini}, L. 2011,
  \mnras, 414, 3566

\bibitem[{{Toomre}(1964)}]{Toomre1964}
{Toomre}, A. 1964, \apj, 139, 1217

\bibitem[{{Turk} {et~al.}(2012){Turk}, {Oishi}, {Abel}, \& {Bryan}}]{Turk2012}
{Turk}, M.~J., {Oishi}, J.~S., {Abel}, T., \& {Bryan}, G.~L. 2012, \apj, 745,
  154

\bibitem[{{Vainshtein} \& {Ruzmaikin}(1971)}]{Vainshtein1971}
{Vainshtein}, S.~I. \& {Ruzmaikin}, A.~A. 1971, \azh, 48, 902

\bibitem[{{Wang} \& {Abel}(2009)}]{WangAbel2009}
{Wang}, P. \& {Abel}, T. 2009, \apj, 696, 96

\bibitem[{{Wang} {et~al.}(2010){Wang}, {Li}, {Abel}, \& {Nakamura}}]{Wang10}
{Wang}, P., {Li}, Z.-Y., {Abel}, T., \& {Nakamura}, F. 2010, \apj, 709, 27

\bibitem[{{Weibel}(1959)}]{Weibel1959}
{Weibel}, E.~S. 1959, Physical Review Letters, 2, 83

\bibitem[{{Widrow} {et~al.}(2012){Widrow}, {Ryu}, {Schleicher}, {Subramanian},
  {Tsagas}, \& {Treumann}}]{Widrow12}
{Widrow}, L.~M., {Ryu}, D., {Schleicher}, D.~R.~G., {et~al.} 2012, \ssr, 166,
  37

\bibitem[{{Yoon} {et~al.}(2014){Yoon}, {Schlickeiser}, \& {Kolberg}}]{Yoon14}
{Yoon}, P.~H., {Schlickeiser}, R., \& {Kolberg}, U. 2014, Physics of Plasmas,
  21, 032109

\bibitem[{{Yoshida} {et~al.}(2008){Yoshida}, {Omukai}, \&
  {Hernquist}}]{Yoshida08}
{Yoshida}, N., {Omukai}, K., \& {Hernquist}, L. 2008, Science, 321, 669

\bibitem[{{Yun} {et~al.}(2001){Yun}, {Reddy}, \& {Condon}}]{Yun01}
{Yun}, M.~S., {Reddy}, N.~A., \& {Condon}, J.~J. 2001, \apj, 554, 803

\end{thebibliography}
 
\newpage

\end{document}